\documentclass[apj]{emulateapj}
\usepackage{apjfonts}

\usepackage{amssymb,latexsym,amsmath}

\def\ltsim{\raise 2pt \hbox {$<$} \kern-1.1em \lower 4pt \hbox {$\sim$}}
\def\gtsim{\raise 2pt \hbox {$>$} \kern-1.1em \lower 4pt \hbox {$\sim$}}

\usepackage{version}

\usepackage {longtable}
\usepackage{rotating}

\usepackage{natbib}
\begin{document}

\shorttitle{A Chandra - VLA investigation of the galaxy cluster RBS 797} 
\shortauthors{Doria et al.}

\title{A Chandra - VLA investigation of the X-ray cavity system and
  radio mini-halo in the galaxy cluster RBS 797}

\author{Alberto Doria\altaffilmark{1}, 
Myriam Gitti\altaffilmark{2,3,4}, 
Stefano Ettori\altaffilmark{3}, 
Fabrizio Brighenti\altaffilmark{2}, 
Paul E. J. Nulsen\altaffilmark{5},
and Brian R. McNamara\altaffilmark{5,6,7}
}

\altaffiltext{1}{ Argelander-Institut f\"{u}r Astronomie, Auf dem H\"{u}gel 71,
D-53121 Bonn, Germany}
\altaffiltext{2}{ Dipartimento di Astronomia, Universit\`a di Bologna, via 
Ranzani 1, Bologna 40127, Italy}
\altaffiltext{3}{ Astronomical Observatory of Bologna - INAF, via
  Ranzani 1, I-40127 Bologna - Italy}
\altaffiltext{4}{ Institute of Radioastronomy - INAF, via
  Gobetti 101, I-40129 Bologna - Italy}
\altaffiltext{5}{ Harvard-Smithsonian Center for Astrophysics, 60
  Garden Street, Cambridge, MA 02138 - USA}
\altaffiltext{6}{ Dept. of Physics \& Astronomy, University of
  Waterloo, 200 University Avenue West, Waterloo, Ontario - Canada N2L
  2G1}
\altaffiltext{7}{ Perimeter Institute for Theoretical Physics, Waterloo, 
Canada}

\begin{abstract}
  We present a study of the cavity system in the galaxy cluster RBS
  797 based on \textit{Chandra} and \textit{VLA} data. RBS 797 (z =
  0.35), is one of the most distant galaxy clusters in which two
  pronounced X-ray cavities have been discovered. The Chandra data
  confirm the presence of a cool core and indicate an higher
  metallicity along the cavity directions. This is likely due to the
  AGN outburst, which lifts cool metal-rich gas from the center along
  the cavities, as seen in other systems. We find indications that the
  cavities are hotter than the surrounding gas.  Moreover, the new
  Chandra images show bright rims contrasting with the deep, X-ray
  deficient cavities. The likely cause is that the expanding 1.4 GHz
  radio lobes have displaced the gas, compressing it into a shell that
  appears as bright cool arms.  Finally we show that the large-scale
  radio emission detected with our VLA observations may be classified
  as a radio mini-halo, powered by the cooling flow (CF), as it nicely
  follows the trend P$_{radio}$ vs.  P$_{CF}$ predicted by the
  re-acceleration model.

\end{abstract}


\maketitle

%%%%%%%%%%%%%%%%%%%%%%%%%%%%%%%%%%%%%%%%%%%%%%%%%%%%%%%%%%%%%%%%%%%%%%%%%%%%%%

\section{Introduction}
\label{intro}

Imaging spectroscopy obtained with the new generation of X-ray
telescopes indicates that the spectra of clusters of galaxies with ``cool
cores'' show no evidence of any cooler phase of the intra-cluster
medium (ICM) below an intermediate temperature of $\sim$1--2 keV.  In
particular, it has also been found that line emission from gas with kT$<$1
keV was either lacking in the spectra of cool core clusters or was weaker than
expected \citep{Peterson_2001, Tamura_2001}. Many possible explanations of
this ``cooling flow problem'' have been suggested in the past decade,
like differential absorption, thermal conduction, intracluster
supernovae, subcluster merging or inhomogeneous metallicity
distributions \citep[e.g.,][for a review]{Peterson-Fabian_2006}.

Heating by a central active galactic nucleus (AGN) in galaxy clusters
has become the most credible scenario to explain the cooling flow
problem \citep[for recent reviews][and references
therein]{Gitti-review_2012, McNamara_Nulsen}.  In this scenario the
central cD galaxies of clusters drive strong jet outflows which
interact with the hot plasma of the ICM inflating lobes of
radio-synchrotron emission. As a consequence, the gas is displaced,
forming X-ray deficient regions, which are thus called ``cavities''.
It is believed that such peculiar features can heat the cluster gas in
various ways.  Furthermore, the cavities are often filled with
relativistic particles and magnetic fields, as indicated by the
observation of radio lobes spatially coincident with the X-ray
cavities \citep[e.g.,][]{Gitti-review_2012, McNamara_Nulsen}.

Studies of individual clusters confirm a strong interplay between the
ICM and the radio sources in their cores, e.g., Perseus
\citep{Boehringer_1993, Churazov_2000, Fabian_2000, Fabian_2006},
A2052 \citep{Blanton_2001, Blanton_2009, Blanton_2011}, MS0735.6+7421
\citep{McNamara_2005, Gitti_2007} and Hydra A \citep{McNamara_2000,
  Nulsen_2005, Wise_2007, Simionescu_2009a, Gitti_2011}. 
Nevertheless, the X-ray luminous, relatively distant galaxy cluster RBS 797 (z =
0.35) seems to be one of the most striking examples of this interaction.

RBS 797 was detected for the first time in the \textit{ROSAT All-Sky
  Survey (RASS)} \citep{Schwope_2000} as the X-ray source RXS
J094713.2+762317, and was observed in the optical band by the
\textit{ROSAT Bright Survey (RBS)} during the optical identification
of all bright (count rate $>$ 0.2 cts s$^{-1}$), high-galactic
latitude ($|b|$ $>$ 30$^\circ$) X-ray sources found in the RASS
\citep{Fischer_RBS, Schwope_2000}.  Early observations with
\textit{Chandra} revealed two pronounced X-ray minima, which are
located on opposite sides of the cluster center \citep{Schindler_2001}
and represent a striking example of a cavity system in a galaxy
cluster. A recent, deeper Chandra image has confirmed the presence of
the cavity system, allowing a more detailed study.  By considering a
complex cavity configuration extended along our line of sight,
\cite{Cavagnolo_2011} provide new estimates of the energy of the
central AGN outburst (up to 6 $\times$ 10$^{60}$ erg) and of the jet
power (up to 6 $\times$ 10$^{45}$ erg s$^{-1}$). Furthermore,
\cite{Cavagnolo_2011} show that the AGN of RBS 797 must be powered
principally by accretion of cold gas, and that its powerful outburst
might be due to the energy released by a maximally spinning black
hole. Structures associated with star formation in the bright central
galaxy have also been found thanks to new \textit{Hubble Space
  Telescope} \citep{Cavagnolo_2011} images.

RBS 797 was observed at radio fequencies with the VLA in the years
2001-2004.  With these observations, performed at different
frequencies and resolutions, it was possible to detect radio emission
on three different scales and orientations, indicating that RBS 797
represents a very peculiar case \citep{Gitti_VLA}.  In particular, the
X-ray cavities were found to be filled with 1.4 GHz radio emission.

Furthermore, diffuse radio emission was detected on a scale of
hundreds of kpc, which is roughly comparable to the size of the
cooling region. This extended radio emission is characterized by
amorphous morphology and a steep spectrum that steepens with distance
from the center.  These characteristics point to a possible
classification of the diffuse radio source as a mini-halo
\citep{Gitti_VLA}.

In this paper we present a new analysis of the recent, deeper
\textit{Chandra} X-ray data. We also present new \textit{VLA}
observations at 1.4 GHz that confirm the presence of large-scale radio
emission. Based on our joint X-ray/radio analysis, we investigate the
physical properties of the ICM and its interaction with the radio
source, and discuss the possibility that RBS 797 hosts a diffuse radio
mini-halo.  In order to be consistent with the literature, we adopt
the cosmology of a $\Lambda$CDM model by the following parameters:
$H_0 = 70 \mbox{ km s}^{-1} \mbox{ Mpc}^{-1}$, and
$\Omega_M=1-\Omega_{\Lambda}=0.3$, the luminosity distance is 1858 Mpc
and 1 arcsec corresponds to 4.8 kpc. Measurement uncertainties are
90\% confidence levels unless stated otherwise.

%%%%%%%%%%%%%%%%%%%%%%%%%%%%%%%%%%%%%%%%%%%%%%%%%%%%%%%%%%%%%%%%%%%%%%%%%%%%%%

\section{Observations and Data Reduction}
\label{observations}

\subsection{Chandra Data}

RBS 797 was observed with the \textit{Chandra Advanced CCD Imaging
  Spectrometer I} (ACIS-I) on October 20th 2000 (ObsID 2202), for a
total exposure of 13.3 ksec \citep{Schindler_2001}.  On July 9th 2007
(ObsID 7902) a new 38.3 ksec were acquired with ACIS-S in imaging
mode, operating at the focal plane temperature of -120$^\circ$ C. The
two Chandra exposures were combined for our analysis, giving a total
of 51.6 ksec of uncleaned expsoure time.

Data were reprocessed with CIAO~4.2 using CALDB~4.2.0 and corrected
for known time-dependent gain problems following techniques similar to
those described in the \textit{Chandra} analysis
threads\footnote{http://cxc.harvard.edu/ciao/threads/index.html}.
Screening of the event files was applied to filter out strong
background flares. Blank-sky background files, filtered in the same
manner as in the RBS 797 image and normalized to the count rate of the
source image in the 10--12 keV band, were used for background
subtraction.  The final exposure time is $\sim$49.6 ksec.  

For both chips, point sources have been identified and removed using
the CIAO task \texttt{WAVDETECT}.  The ``holes'' left by the discarded
point sources are re-filled, for imaging purposes only, by values
interpolated from surrounding background regions using the task
\texttt{dmfilth}.
% with the detection threshold set to the default value of 10$^{-6}$.
Images, instrument maps, and exposure maps were created in the
0.5--7.0 keV band.
% using the CIAO tool \texttt{dmcopy} and \texttt{dmextract}.
Data with energies above 7.0 keV and below 0.5 keV were excluded in
order to prevent background contamination and uncertainties in the
ACIS calibration, respectively.

\begin{figure*}[ht]
  \centerline{
	\includegraphics[width=0.5\textwidth]{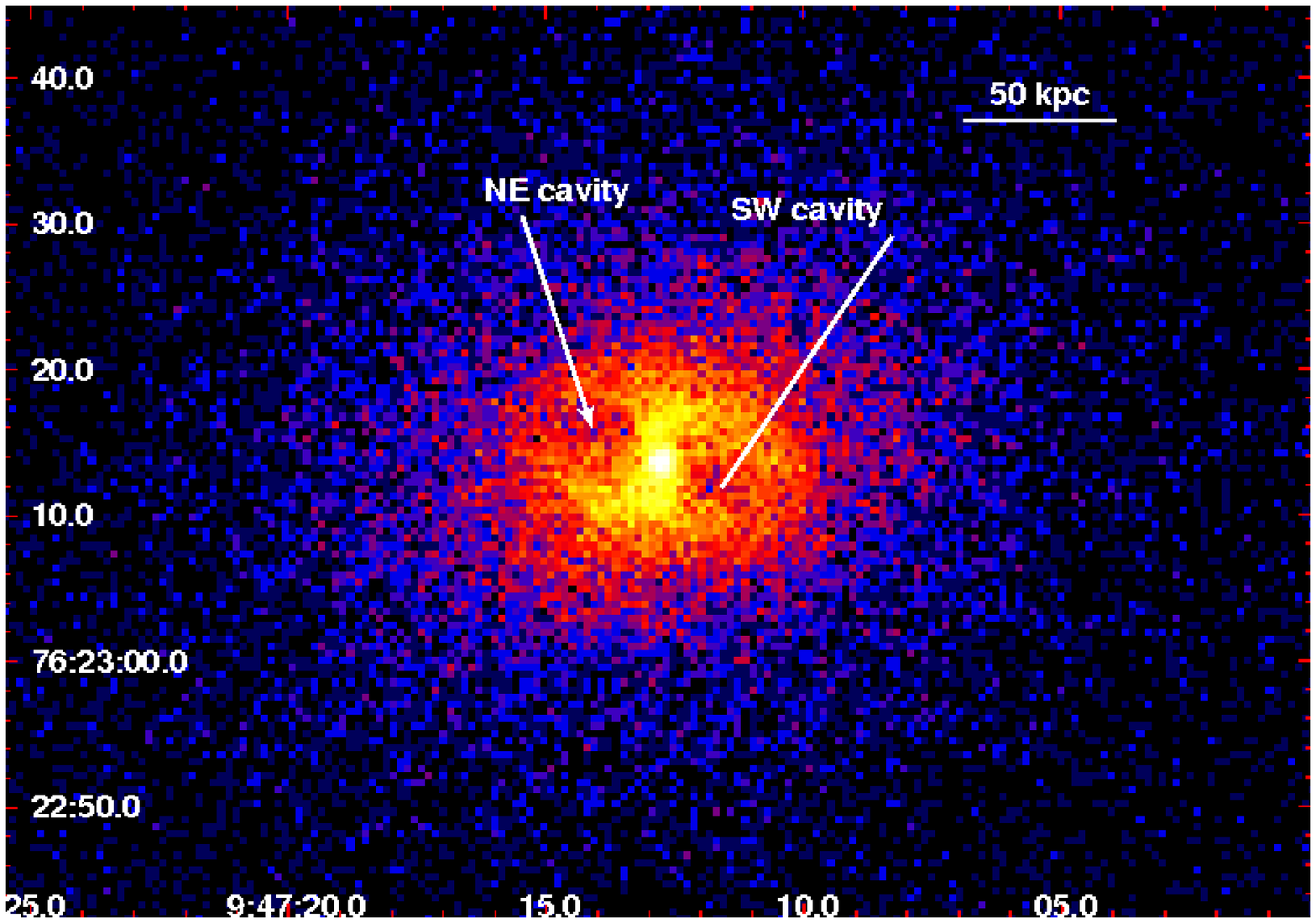}
	\includegraphics[width=0.5\textwidth]{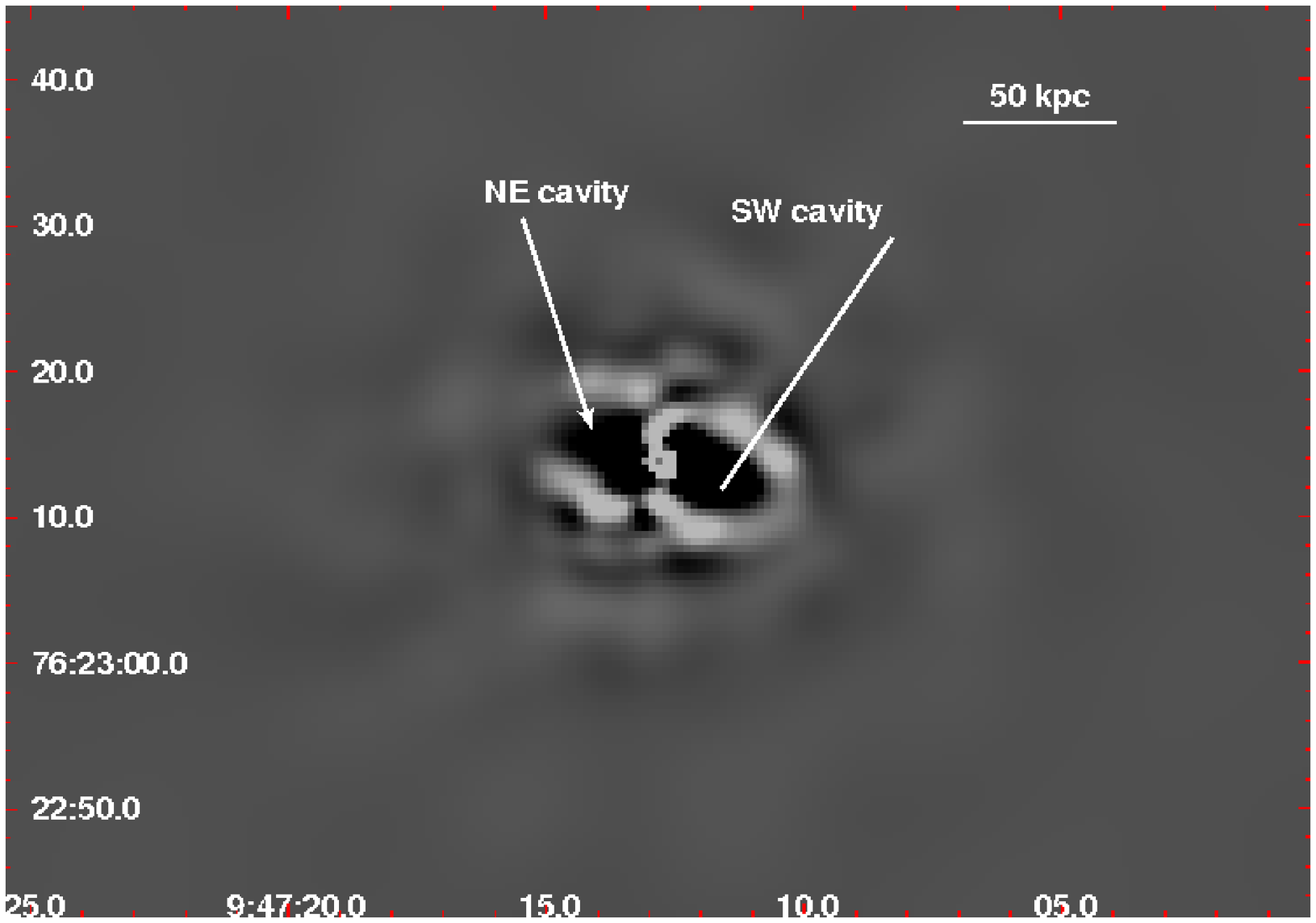}
}
\caption{Raw (left panel) and unsharp masked (right panel) Chandra ACIS-S
images of RBS 797 in the 0.5--7.0 keV energy band. The unsharp masked image is
obtained by subtracting a strongly (S/N=10) smoothed image from a lightly
(S/N=3) smoothed image.  In these images and in all the following, north is up
and east is left.}
	\label{fig:raw+unsharp}
\end{figure*}

%%%%%%%%%%%%%%%%%%%%%%%%%%%%%%%%%%%%%%%%%%%%%%%%%%%%%%%%%%%%%%%%%%%%%%%%%%%%%%

\subsection{VLA Data}
\label{vla.sec}

The existing \textit{Very Large Array}\footnote{The \textit{Very Large
    Array} (VLA) is a facility of the \textit{National Radio
    As\-tro\-no\-my Observatory} (NRAO).  The NRAO is a facility of
  the National Science Foundation, operated under cooperative
  agreement by Associated Universities, Inc.} data for the radio
source RBS 797 at 300, 1400, 4800 and 8400 MHz have already been presented
and discussed in \citet{Gitti_VLA}, \citet{Birzan_2008} and
\citet{Cavagnolo_2011}. Here we present new observations at 1.4 GHz
performed in December 2006 with the VLA in C array, a configuration
that was not used in previous observations.  The data were acquired in
channels centered at 1465 and 1385 MHz, with 50 MHz bandwidth, for a
total integration time of 5 hours. In these observations the source 3C
48 (0137$+$331) is used as the primary flux density calibrator, while
the source 1044+809 is used as secondary phase calibrator.  We have
also performed a re-analysis of the archival 1.4 GHz data in A- and B-
array, and then have added them together with the new 1.4 GHz data in
C-array to produce a total, combined 1.4 GHz image.

Data reduction was done using the NRAO AIPS (Astronomical Image
Processing System) package. Accurate editing of the UV data was
applied to identify and remove bad data. Images were produced by
following the standard procedures: Calibration, Fourier-Transform,
Clean and Restore. Self-calibration was applied to remove residual
phase variations. The final images, produced using the AIPS task
IMAGR, show the contours of the total intensity.

%%%%%%%%%%%%%%%%%%%%%%%%%%%%%%%%%%%%%%%%%%%%%%%%%%%%%%%%%%%%%%%%%%%%%%%%%%%%%%

\section{Results}

\subsection{X-ray Morphology and Surface Brightness Profile}
\label{subsec:morphology}

The pronounced central surface brightness peak (discussed below in
this Section) and the presence of strong O[II]$\lambda$3727 line, and
weak, narrow H$\beta$ and O[III]$\lambda\lambda$4959/5007 emission
lines \citep{Schindler_2001} indicate the presence of a central AGN
recognized by \cite{Schindler_2001}. In order to avoid contamination
by the central source, we exclude it in the spectral analysis (Section
\ref{sec:Spectral Analysis}).

The raw ACIS-S image of RBS 797 in the 0.5--7 keV energy band is shown
in the left panel of Figure \ref{fig:raw+unsharp}.  We confirm the
presence of two central X-ray deficient lobes already discovered with
the existing \textit{Chandra} observations of RBS 797
\citep{Schindler_2001, Cavagnolo_2011}. These two striking features
are $\sim$10 arcsec apart in a NE - SW orientation (Figure
\ref{fig:raw+unsharp}). The physical properties of the cavities are
discussed in Section \ref{subsec:energetics}. The cavity system in RBS
797 is surrounded by bright rims. One possible explanation for the
presence of these strong features is that the expansion of the 1.4 GHz
radio lobes displaced the ICM formerly in the cavities, compressing it
to form the bright rims that now surround the cavities (as seen in the
right panel of Figure \ref{fig:raw+unsharp}).  In particular, the rims
of the SW cavity appear to have a stronger contrast than the NE cavity
rims, almost totally enclosing the SW cavity, while the NE rims seem
to be more open and less bright.  Beyond the cavities and rims, the
cluster morphology is elliptical in shape.

These details are more evident in the \textit{Chandra} unsharp masked
image (Figure \ref{fig:raw+unsharp}, right panel), which is created
with an image reprocessing tecnhique aimed at emphasizing possible
substructures and faint features. We also note in the unsharp masked
image that the cavity system appears to be surrounded by two edges,
with the inner edge externding $\sim$8$''$--10$''$ from the core. An
outer edge is also visible at $\sim$14$''$--17$''$.  These features
appear as edges in the surface brightness profile at $\sim$8$''$ and
$\sim$20$''$ (see below). Taking into account these peculiarities, RBS
797 appears to be very similar to NGC 5813 \citep{Randall_2011} in
which are observed distinct cavity systems due to multiple AGN
outbursts.  Also in RBS 797, as in NGC 5813, the edges seen in the
\textit{Chandra} X-ray image may be identified as weak shocks (for a
further discussion about shocks see Section 3.3.3 in
\citealp{Cavagnolo_2011}).  These shocks could be produced by distinct
AGN outbursts, with the inner edge associated to the NE-SW inner
cavities, and the outer edge associated to possible outer cavities.

%%%% SURFACE BRIGHTNESS PROFILES %%%%%%

We also performed a study of the surface brightness profile of RBS 797
by means of the {\it Chandra} ACIS-S data.  An azimuthally averaged
surface brightess profile was computed over the range 0.5--2.0 keV
with the task \texttt{dmextract} by subtracting the background from
the point-source-removed image and then dividing by the exposure map.

\begin{figure}[b]
\includegraphics[width=0.48\textwidth]{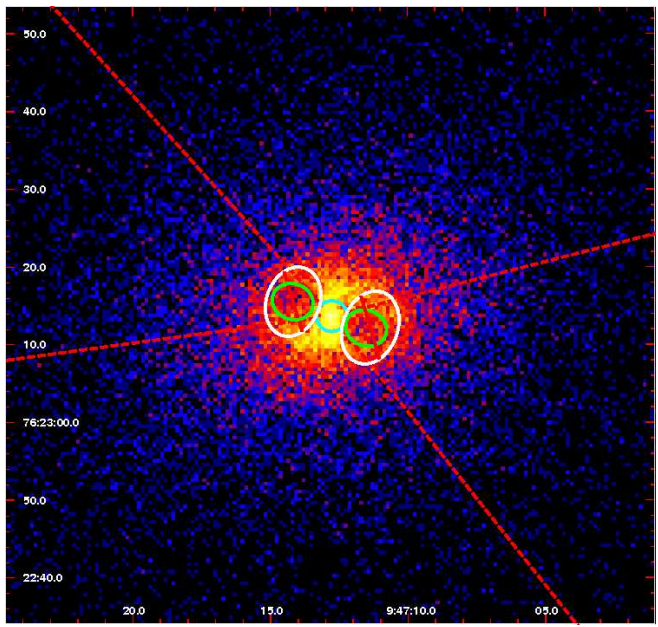}

\caption{Raw ACIS-S image of RBS 797 in the 0.5--2.0 keV energy
  interval. The surface brightness profile of the ``undisturbed'' ICM
  was obtained by considering 360$^\circ$~annular bins of 1$''$ width
  and excluding the white elliptical regions which contain the
  cavities (green ellipses) and the rims.  The red dashed lines
  delimit the two sectors considered for the surface brightness
  profile calculated along the cavity directions. See text for
  details.}
			\label{fig:SB_unp}
\end{figure}

\begin{figure}[b]	 
\includegraphics[width=0.48\textwidth]{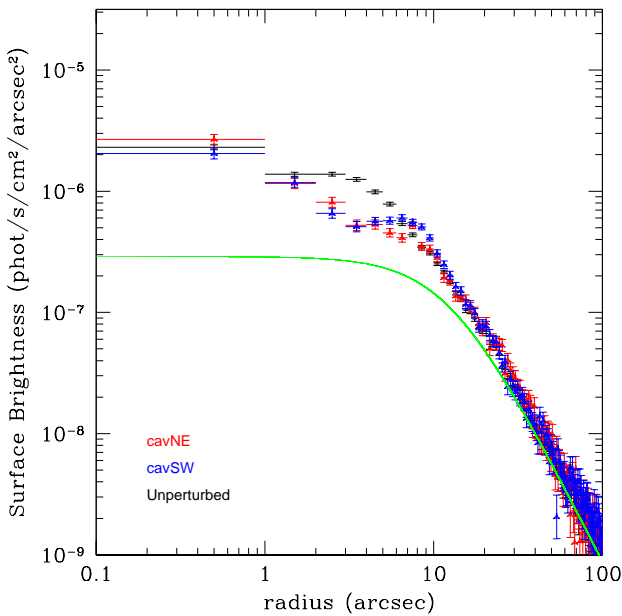}

\caption{Background subtracted, azimuthally-averaged 0.5--2.0 keV
  radial surface brightness profile of the undisturbed cluster (black
  crosses), compared with the surface brightness profile along the NE
  (red triangles) and SW (blue triangles) cavity regions (see Figure
  \ref{fig:SB_unp}). Errorbars indicate uncertainties at 1$\sigma$
  confidence.
  Overlaid in green is the fit with a single $\beta$-model ($\chi^{2}_{red}$ =
  0.89) of the surface brightness profile of the undisturbed cluster. It has
  been performed in the 25$''$--110$''$ radial range and extrapolated to the 
  center.
}
			\label{fig:SB}
\end{figure}

In particular, we considered 360$^\circ$~annular bins of 1$''$ width,
starting from 2$''$ up to 150$''$. Since there is clear evidence of
ICM / AGN interaction in the central region of RBS 797, we excluded
the cavity system and the rims (see Figure \ref{fig:SB_unp}) in order
to get a profile of the so-called ``undisturbed'' ICM. This surface
brightness profile was fitted with a single $\beta$-model
\citep*{Cavaliere_beta} using the tool \texttt{SHERPA} with the
$\chi^{2}$ statistics with Gehrels' variance.
A fit over the external 25$''$--110$''$ interval, which excludes
the whole cooling region (green line in Figure \ref{fig:SB}), shows
evidence of a central excess of the observed profile with respect to
the $\beta$-model (Figure \ref{fig:SB}). This evident deviation from
the fit prediction is a strong indication of the presence of a cool
core in RBS 797, as seen in many other cool core clusters
\citep{Fabian_1994}. We then fitted the surface brightness profile
of the undisturbed cluster with a double $\beta$-model, finding that
it is able to provide a better description of the entire profile
($\chi^{2}_{red}$ $\sim$ 0.96 for 142 d.o.f.). The best-fit values for
the double $\beta$-model (assuming a common $\beta$ value) are: core
radii r$_c$(1) = 3$''$.47 $\pm$ 0$''$.28 and r$_c$(2) = 12$''$.89
$\pm$ 1$''$.13, central surface brightness S$_{0}$(1) = 1.15 $\pm$
0.06 $\times$ 10$^{-6}$ and S$_{0}$(2) = 2.21 $\pm$ 0.50 $\times$
10$^{-7}$, and slope parameters $\beta$(1) = 0.6044 $\pm$ 0.0543 and
$\beta$(2) = 0.6332 $\pm$ 0.0194.

%%%% DROPS IN SURFACE BRIGHTNESS %%%%%%%%

For a comparison of the surface brightness profile along the cavities
with the azimuthally-averaged profile, we considered two sectors along
the cavity directions, chosen to be tangent to the cavities (Figure
\ref{fig:SB_unp}, in red).  Figure \ref{fig:SB} also shows the two
profiles along the cavities compared separately with the
azimuthally-averaged profile. We found a clear drop in the radial
range between 2$''$ and 7$''$ in both the surface brightness profiles
along the cavities compared to the azimuthally-averaged profile. These
distances clearly correspond to the two X-ray deficient lobes, thus
giving us a very good indication for the estimate of the cavity
dimensions (see Section \ref{subsec:energetics}).

%%%%%%%%%%%%%%%%%%%%%%%%%%%%%%%%%%%%%%%%%%%%%%%%%%%%%%%%%%%%%%%%%%%%%%%%%%%%%%

\subsection{Radio Properties}
\label{results-radio.sec}

RBS 797 is known to possess a unique radio source that exhibits large
changes in orientation with scale.  The subarsec resolution radio
image first presented by \citet{Gitti_VLA} shows the details of the
innermost 4.8 GHz radio jets, which clearly point in a north-south
direction.  Remarkably, these inner jets are almost {\it
  perpendicular} to the axis of the 1.4 GHz emission detected at
$\sim$1 arcsec resolution (A-array), which is elongated in the
northeast-southwest direction filling the X-ray cavities \citep[see
Fig. 4 of][]{Gitti_VLA}. The $\sim$4 arcsec resolution VLA image at
1.4 GHz (B-array) shows large-scale radio emission
with amorphous morphology, slightly extended in the north-south
direction \citep[see Figure 2 of][]{Gitti_VLA}.

\begin{figure*}[t]
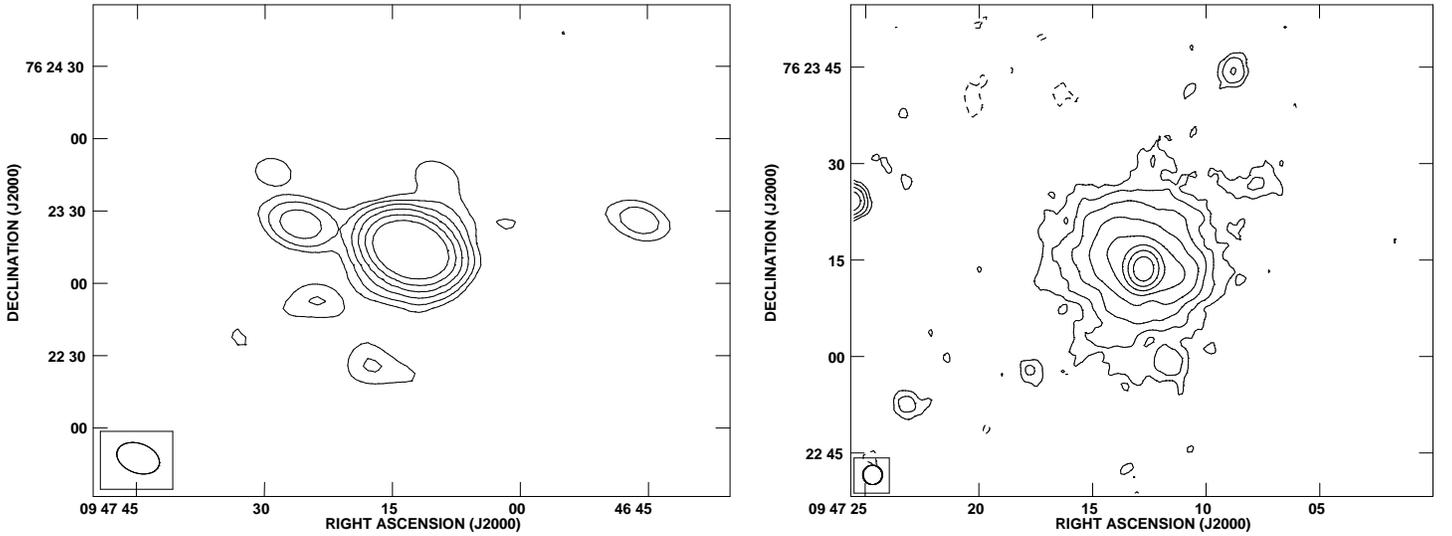

\vspace{0cm}
\centerline{
\includegraphics[width=8.3cm, angle=-90]{doria_fig4a.ps}
\includegraphics[width=8.3cm, angle=-90]{doria_fig4b.ps}
}
\vspace{0cm}
\caption{\label{vla-new.fig} {\it Left:} 1.4 GHz VLA (C-array) map of
  RBS 797 at a resolution of $18''.3 \times 12''.2$ (the beam is shown
  in the lower-left corner).  The r.m.s. noise is 0.03 mJy/beam. The
  first contour corresponds to 3$\sigma$, the ratio between two
  consecutive contours being 2. {\it Right:} Total 1.4 GHz VLA map of
  RBS 797 at a resolution of $3''$ (the beam is shown in the
  lower-left corner) obtained by combining the observations in A-, B-,
  and C- array.  The r.m.s. noise is 0.01 mJy/beam. The contour levels
  are $-$0.035 (dashed), 0.035, 0.070, 0.140, 0.280, 0.560, 1.120,
  2.240, 4.480 mJy/beam.}
\end{figure*}

Figure \ref{vla-new.fig} (left panel) shows the radio map of RBS 797
observed at 1.4 GHz with the VLA in C-array configuration, with a
restoring beam of $18''.3 \times 12''.2$. At this low resolution the
source is barely resolved, showing an amorphous morphology apparently
extending to the north (an unrelated source is present to the east).
The source has a total flux density of $\simeq 23.0 \pm 0.3$ mJy.
%\footnote{1 Jy = $10^{-26} \, {\rm W} \, {\rm Hz}^{-1} {\rm m}^{-2}$}.
In order to fully exploit the relative advantages in terms of angular 
resolution and sensitivity of the three VLA configurations used for          
the observations at 1.4 GHz, we have combined this new set of data         
taken with the C-array with the archival data from the A- and B-          
array. The total 1.4 GHz radio map, with a circular restoring beam of
$3''$, is shown in Figure \ref{vla-new.fig} (right panel). This 
combined image is able to reveal the morphology of the radio source in 
the central $\sim 10$ kpc, showing its elongation in the cavity 
direction, without losing flux at the larger scales. In particular,
the radio emission is detected out to $\sim 90$ kpc. The total flux 
density is $\simeq 24.0 \pm 0.3$ mJy.

In order to estimate the flux density of the diffuse radio emission it
is necessary to subtract the emission contributed by the central
nuclear source. One way is to consider the high-resolution map. The
central component imaged by the A-array observations \citep[see Figure
2 of][]{Gitti_VLA} is found to consist of a resolved core elongated in
the northeast-southwest direction, plus a jet-like feature emanating
from the center to the west. The total flux density coming from the
central region (inner jets included) was measured by \citet{Gitti_VLA}
as $\simeq 12.1 \pm 0.1$ mJy. Another way is to estimate the
contribution of the central emission directly from the new combined
image presented here (Figure \ref{vla-new.fig}, right panel). In this
map, the flux density of the unresolved core is $\simeq 12.9 \pm 0.1$
mJy.  The slight difference between the two estimates may be due to
the contribution of the short baselines in the combined image with
respect to the image in A-array only.  To be conservative, we decided
to adopt the mean of the two flux densities measured above.
We thus estimate that the discrete radio source located at the cluster
center has an 1.4 GHz flux density of $\simeq 12.5 \pm 0.5$ mJy, where
the errors include the mi\-ni\-mum and maximum variation of such an
estimate based on our analysis. By subtracting this from the total
flux density of the 1.4 GHz emission we measure a diffuse radio
emission of $\simeq 11.5 \pm 0.6$ mJy.  The more drastic approach of
subtracting the total flux density measured in A-array \citep[$17.9
\pm 0.2$ mJy,][]{Gitti_VLA}, i.e. the entire radio source filling the
X-ray cavities, would still leave a residual $\simeq 6.1 \pm 0.4$ mJy
of diffuse radio emission.  The variety of methods discussed here
demonstrates the difficulty of disentangling the relative contribution
of the diffuse emission and of the discrete sources in the total
observed radio emission.  The nature of the large-scale radio
emission, and its possible origin, will be discussed in section
\ref{minihalo.sec}.

%%%%%%%%%%%%%%%%%%%%%%%%%%%%%%%%%%%%%%%%%%%%%%%%%%%%%%%%%%%%%%%%%%%%%%%%%%%%%%

\section{Chandra Spectral Analysis} 
\label{sec:Spectral Analysis} 

In each region of interest, a single spectrum was extracted with the
CIAO task \texttt{specextract} and then grouped to give at least 25
counts in each bin. Only counts deriving from energy bands between 0.5
and 7.0 keV were included.  Spectra have been extracted separately
from ACIS-S3 and ACIS-I3 observations with independent response
matrices and then fitted jointly with the tool \texttt{XSPEC} version
\texttt{12.5.1n}, using a \texttt{wabs*apec} model, approximating a
collisionally-ionized diffuse gas subject to absorption. The hydrogen
column density was fixed at the Galactic value \citep[N$_{H}$ = 2.28
$\times$ 10$^{20}$ cm$^{-2}$,][]{Kalberla_2005} and the redshift was
frozen at \textit{z} = 0.35. The metallicity parameter was always
allowed to be free and was measured relative to the abundances of
\citet{Anders-Grevesse_1989}.  The central circle with radius of
1$''$.5 (7.2 kpc) contains the central X-ray AGN. In order to avoid
contamination with the thermal emission of the ICM, this region was
not considered in the spectral fitting.  In addition, regions of the
cluster lying close to the outer edges of the I3 CCD were excluded,
restricting the analysis to the first 125$''$ (600 kpc).  The global
cluster temperature, measured in the annulus 1$''$.5-125$''$, is $kT =
6.73^{+0.26}_{-0.20}$ keV.

%%%%%%%%%%%%%%%%%%%%%%%%%%%%%%%%%%%%%%%%%%%%%%%%%%%%%%%%%%%%%%%%%%%%%%%%%%%%%%

\subsection{Azimuthally Averaged Profiles: Temperature and Metallicity}
\label{subsec:Azimuthally Averaged Profiles}

\begin{figure*}[htb]
\includegraphics[width=0.33\textwidth]{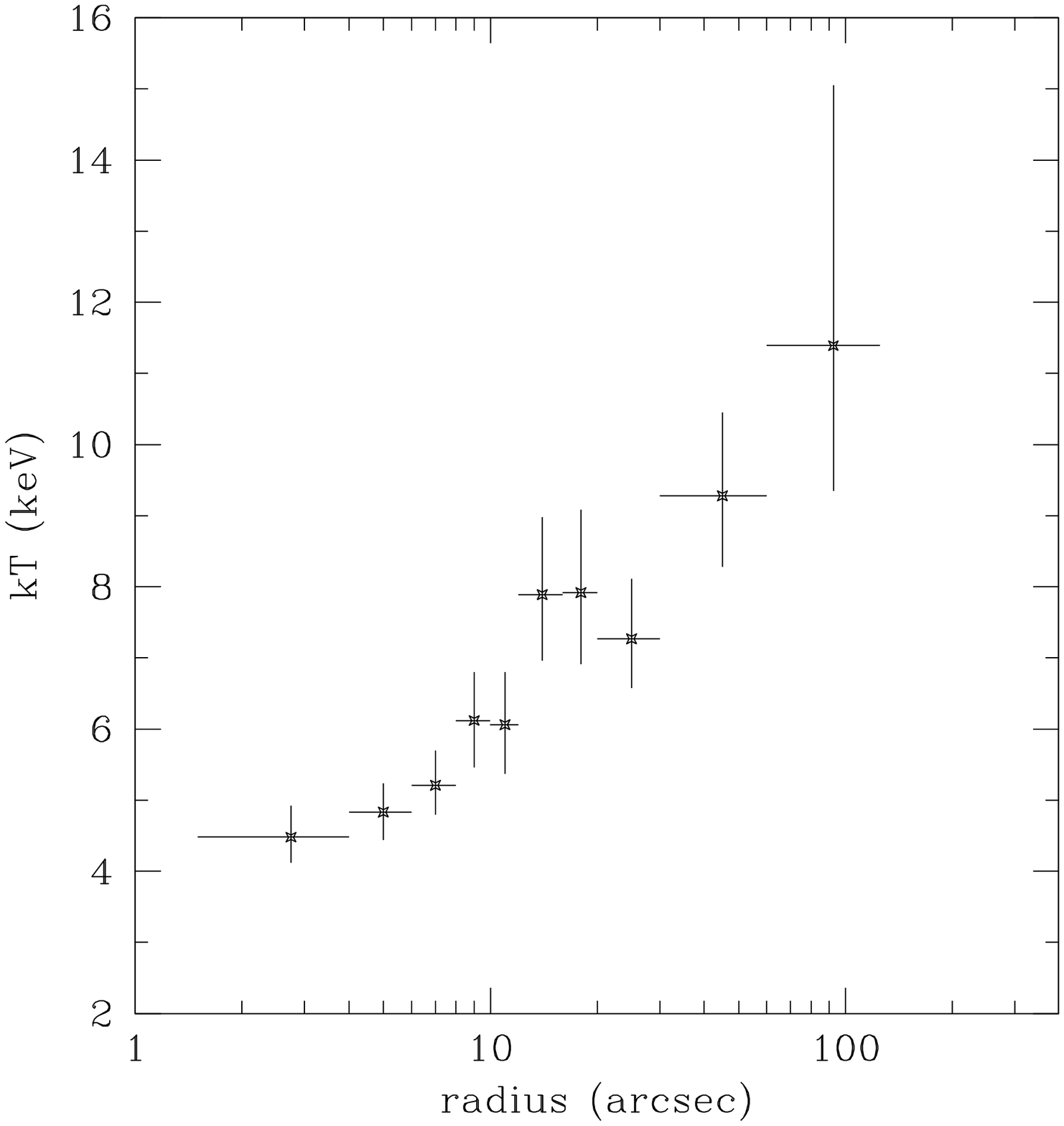}
\includegraphics[width=0.33\textwidth]{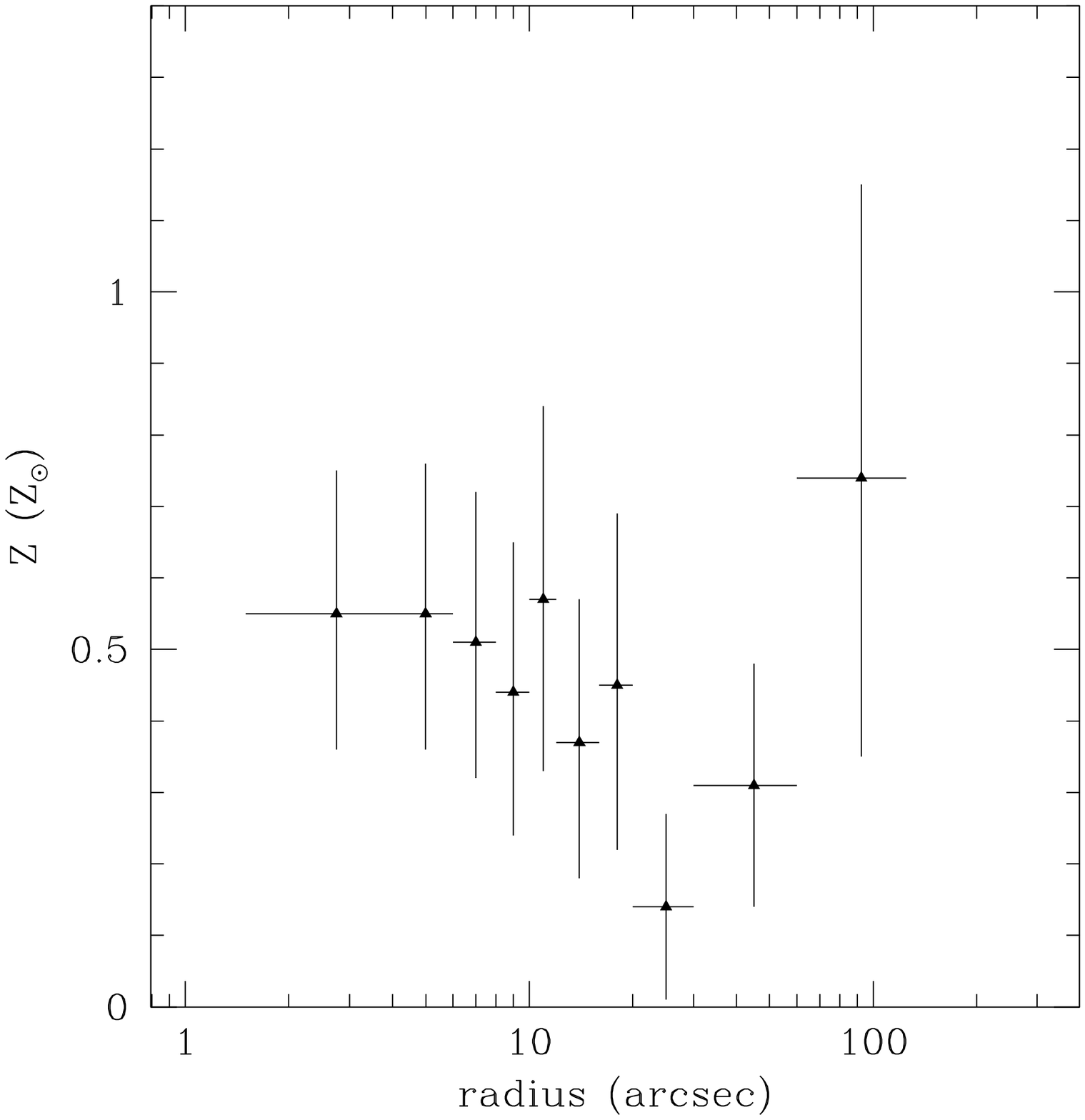}
\includegraphics[width=0.33\textwidth]{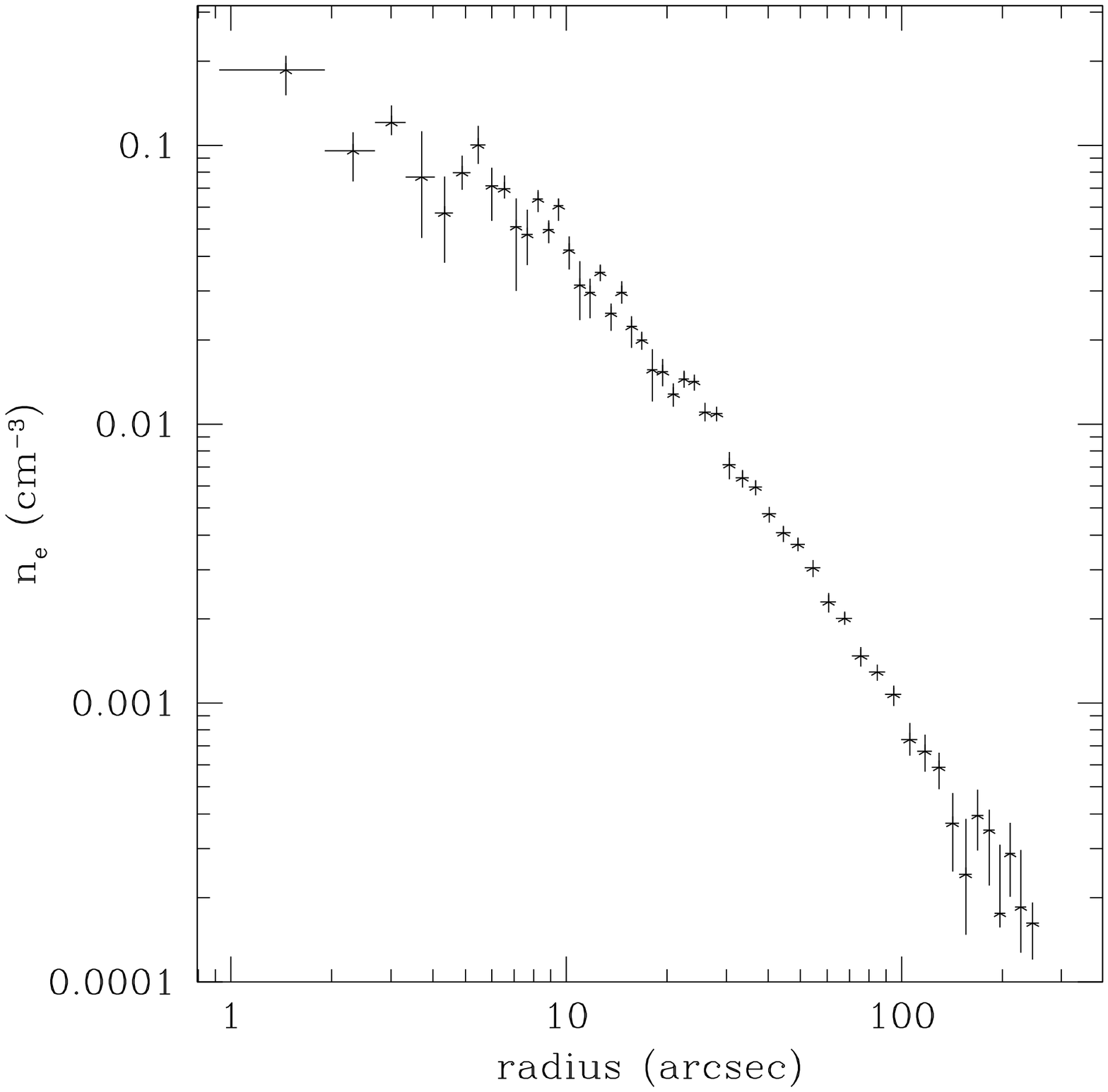}
\caption{\textit{Left panel:} azimuthally-averaged, projected
  temperature profile obtained by fitting a \texttt{wabs*apec}
  model. The errorbars on the x-axis indicate the radius of the
  annulus considered for each bin. Neither the central AGN, nor the
  very external regions of the cluster are considered.
  \textit{Central panel:} Same as left panel, for the metallicity
  profile. \textit{Right panel:} ICM density profile obtained from the
  0.5--2.0 keV surface brightness profile (see text for
  details). Vertical bars indicate 1$\sigma$ errors.}
			\label{fig:rad_met_temp}
\end{figure*}

A projected radial profile was computed using ten concentic annuli
centered on the cluster core. The radii were selected in order to
obtain sufficient counts to constrain the temperature to at least
$\sim$20\% accuracy in each annulus (except the outermost annulus,
see Table \ref{tab:temp_profile} for details).  The only parameters
free to vary in this \texttt{wabs*apec} model are the metallicity,
\textit{Z}, and the temperature \textit{kT}. Spectra were fitted in
the range 0.5--7.0 keV. The best-fitting parameters and re\-la\-ti\-ve
errors are listed in Table \ref{tab:temp_profile}. The measured
temperature and metallicity profiles are shown in Figure
\ref{fig:rad_met_temp} (left and central panel, respectively).  The
temperature values in the outer two annuli are surprisingly high and
do not show the decline in the outskirt of the cluster.  To test the
reliability of these measurements, we adopted as a different
background spectrum that extracted in a circular annulus in the outer
regions of the two CCDs, finding similar results.

\begin{deluxetable*}{ccccccc}[t]
\tabletypesize{\scriptsize}
\tablewidth{0pt}
\tablecaption{ Azimuthally Averaged Spectral Properties
\label{tab:temp_profile}
 }
\tablehead{
\colhead{r1 - r2} & \multicolumn{2}{c}{r} & \colhead{Source} & {kT} &
{Z} & {$\chi^{2}$/dof}\\
\colhead{($''$)} & \colhead{($''$)} & \colhead{(kpc)} & {Counts} & {(keV)} &
{(Z$_{\odot}$)}
}
\startdata
1.5 - 4	&	2.75	&	13.2	&	3856	&
4.48$_{-0.36~(0.24)}^{+0.44~(0.26)}$	&	0.55$_{-0.19~(0.11)}^{+0.20~
(0.12)}$ & 107.2/115 \\
[+1mm]
		4 - 6	&	5	&	24	&	4118	&
4.83$_{-0.39~(0.24)}^{+0.41~(0.25)}$	&	0.55$_{-0.19~(0.12)}^{+0.21~
(0.12)}$ & 119.2/120 \\
[+1mm]
		6 - 8	&	7	&	33.6	&	4120	&
5.21$_{-0.41~(0.25)}^{+0.49~(0.27)}$ 	&	0.51$_{-0.19~(0.12)}^{+0.21~
(0.13)}$ & 129.2/123\\
[+1mm]
		8 - 10	&	9	&	43.2	&	3512	&
6.12$_{-0.66~(0.40)}^{+0.68~(0.41)}$	&	0.44$_{-0.20~(0.12)}^{+0.21~
(0.13)}$ & 130.0/108\\
[+1mm]
		10 - 12	&	11	&	52.8	&	2867	&
6.06$_{-0.69~(0.44)}^{+0.74~(0.44)}$ 	&	0.57$_{-0.24~(0.15)}^{+0.27~
(0.16)}$ & 84.4/86\\ [+1mm]
		12 - 16	&	14	&	67.2	&	4619	&
7.89$_{-0.93~(0.57)}^{+1.09~(0.58)}$	&	0.37$_{-0.19~(0.12)}^{+0.20~
(0.12)}$ & 144.2/137 \\
[+1mm]
		16 - 20	&	18	&	86.4	&	3591	&
7.92$_{-1.01~(0.61)}^{+1.16~(0.63)}$	&	0.45$_{-0.23~(0.14)}^{+0.24~
(0.14)}$ & 99.1/111 \\
[+1mm]
		20 - 30	&	25	&	120	&	6099	&
7.27$_{-0.69~(0.47)}^{+0.84~(0.50)}$	&	0.14$_{-0.13~(0.08)}^{+0.13~
(0.08)}$ & 141.5/172 \\
[+1mm]
		30 - 60	&	45	&	216	&	7607	&
9.28$_{-1.00~(0.69)}^{+1.17~(0.68)}$	&	0.31$_{-0.17~(0.10)}^{+0.17~
(0.10)}$ & 221.0/210 \\
[+1mm]
		60 - 125	&	92.5	&	444	&	4897
&	11.39$_{-2.04~(1.35)}^{+3.66~(1.90)}$	&	0.74$_{-0.39~
(0.23)}^{+0.41~(0.24)}$ &
227.9/215 \\[-2.5mm]
\enddata
\tablecomments{Results of the spectral fitting of the
  360$^\circ$~annular regions in the 0.5--7.0 keV energy range
  obtained using the \texttt{wabs*apec} model with the absorbing
  column density fixed to 2.28$\times$10$^{20}$. Temperature (in keV),
  abundances (in fraction of solar value) and normalizations are left
  as free parameters. The central 1$''$.5 including the X-ray AGN are
  excluded from the analysis. The first column lists the delimiting
  radii for each annulus, while the second and the third list the
  central value for each bin.  Errors are at 90\% of confidence level,
  while those enclosed in brackets are at 1$\sigma$.}
\end{deluxetable*}

There is indication of a discontinuity in the projected temperature
profile (left panel of Figure \ref{fig:rad_met_temp}) between the bins
10$''$--12$''$ and 12$''$--16$''$, which correspond to the region of
the rims just outside the cavities.  The limited statistics and the
edge of the ACIS-I chip do not allow us to measure the behaviour of
the temperature profile beyond 125$''$.  Therefore, we are not able to
observe the decline of the temperature typically observed at larger
radii in relaxed clusters \citep{Vikhlinin_temp_prof}.

The metallicity profile of RBS 797 (central panel of figure 2) shows a
hint of gradient, with an average abundance of $\sim$0.4
Z$_{\odot}$. A more detailed analysis of the azimuthally averaged
metallicity profile is not possible due to the large uncertainties.

\subsection{Deprojection Analysis} 
\label{subsec:Deprojection Analysis}

Because of the projection effects, the spectral properties at any
point in the cluster are the emission-weighted superposition of
radiation originating at all points along the line of sight through
the cluster.  To correct for this effect, we performed a deprojection
analysis by adopting the \texttt{XSPEC projct} model.
We thus fit the spectra extracted in the same ten
360$^\circ$~concentric annular regions adopted for the projected
temperature profile (see Section \ref{subsec:Azimuthally Averaged
  Profiles}) with a \texttt{projct*wabs*apec} model.
Temperature, abundance and normalization values are free to vary,
while the density column and the redshift are frozen at the same
values adopted for the \texttt{wabs*apec} model.

Results of the deprojection analysis are listed in Table
\ref{tab:temp_profile_dpr}.
We note that the deprojection analysis increases the errorbars without
producing significant changes in the global profile shapes. In
particular the metallicity measurement is strongly limited by poor
statistics. Therefore in the following discussion we adopt the
projected values of metallicity.

We also derived the density profile by deprojecting the surface
brightness profile \citep[see][for details of the
method]{Ettori_2002}, obtaining a typical profile of a cool core
cluster (Figure \ref{fig:rad_met_temp}, right panel).  This is in
agreement with the density profile derived from the spectral
deprojection analysis (Table \ref{tab:temp_profile_dpr}).

\begin{deluxetable*}{ccccccc}[t]
\tabletypesize{\scriptsize}
\tablewidth{0pt}
\tablecaption{ Deprojected Spectral Analysis
\label{tab:temp_profile_dpr}
 }
\tablehead{
\colhead{r1 - r2} & \multicolumn{2}{c}{r} & \colhead{kT} & \colhead{Z}	&
\colhead{norm}	&	\colhead{n$_{e}$}\\[+1mm]
\colhead{($''$)} & \colhead{($''$)} & \colhead{(kpc)} &
\colhead{(keV)} & \colhead{(Z$_{\odot}$)}	&
($\times$10$^{-4}$)	&	($\times$10$^{-3}$ cm$^{-3}$)
}
\startdata
1.5 - 4	&	2.75	&	13.2	&
3.77$_{-0.74~(0.49)}^{+1.61~(0.43)}$	&
0.70$_{-0.70~(0.45)}^{+0.90~(0.48)}$	&
3.32$_{-0.76~(0.45)}^{+0.74~(0.46)}$	&
105.38$_{-12.06~(7.14)}^{+11.74~(7.30)}$\\[+1mm]
		4 - 6	&	5	&	24	&
4.61$_{-1.06~(0.72)}^{+2.19~(1.08)}$	&
0.51$_{-0.51~(0.43)}^{+0.98~(0.56)}$	&
5.46$_{-1.21~(0.73)}^{+0.90~(0.56)}$	&
85.35$_{-9.46~(5.71)}^{+7.03~(4.38)}$\\[+1mm]
		6 - 8	&	7	&	33.6	&
4.32$_{-0.64~(0.37)}^{+1.32~(0.72)}$ 	&
0.73$_{-0.62~(0.39)}^{+0.72~(0.40)}$	&
6.74$_{-1.16~(0.68)}^{+1.21~(0.74)}$	&
67.95$_{-5.85~(3.43)}^{+6.10~(3.73)}$\\[+1mm]
		8 - 10	&	9	&	43.2	&
6.27$_{-1.49~(0.97)}^{+2.70~(1.53)}$	&
0.18$_{-0.18~(0.18)}^{+0.59~(0.34)}$	&
6.96$_{-0.94~(0.37)}^{+0.84~(0.55)}$	&
53.78$_{-3.63~(1.43)}^{+3.24~(2.12)}$\\[+1mm]
		10 - 12	&	11	&	52.8	&
4.44$_{-0.73~(0.47)}^{+1.21~(0.68)}$	&
0.88$_{-0.58~(0.34)}^{+0.59~(0.32)}$	&
5.63$_{-1.03~(0.62)}^{+1.02~(0.57)}$	&
39.60$_{-3.62~(2.18)}^{+3.59~(2.00)}$\\[+1mm]
		12 - 16	&	14	&	67.2	&
8.26$_{-2.01~(1.36)}^{+2.10~(1.55)}$	&
0.26$_{-0.26~(0.25)}^{+0.36~(0.23)}$	&
8.67$_{-0.84~(0.44)}^{+0.84~(0.39)}$	& 
27.25$_{-1.32~(0.69)}^{+1.32~(0.61)}$\\[+1mm]
		16 - 20	&	18	&	86.4	&
7.54$_{-1.34~(0.91)}^{+1.84~(1.16)}$	&
0.83$_{-0.48~(0.32)}^{+0.46~(0.30)}$	&
7.31$_{-0.83~(0.50)}^{+0.81~(0.48)}$	&
19.49$_{-1.11~(0.67)}^{+1.08~(0.64)}$\\[+1mm]
		20 - 30	&	25	&	120	&
7.21$_{-0.82~(0.53)}^{+1.12~(0.68)}$	&
0.08$_{-0.08~(0.08)}^{+0.17~(0.10)}$	&
15.90$_{-0.72~(0.38)}^{+0.66~(0.43)}$	&
13.03$_{-0.29~(0.16)}^{+0.27~(0.18)}$\\[+1mm]
		30 - 60	&	45	&	216	&
9.03$_{-1.08~(0.71)}^{+1.41~(0.85)}$	&
0.26$_{-0.19~(0.12)}^{+0.19~(0.10)}$	&
16.76$_{-0.72~(0.43)}^{+0.73~(0.44)}$	&
4.24$_{-0.09~(0.05)}^{+0.09~(0.06)}$\\[+1mm]
		60 - 125	&	92.5	&	444	&
12.33$_{-2.40~(1.74)}^{+3.85~(2.36)}$	&
0.74$_{-0.39~(1.17)}^{+0.36~(0.20)}$	&
11.77$_{-0.88~(0.55)}^{+1.04~(0.61)}$	&
1.17$_{-0.04~(0.03)}^{+0.05~(0.03)}$ \\[-2.5mm]
\enddata
\tablecomments{Results of the deprojected analysis of the
  360$^\circ$~annular regions in the 0.5--7.0 keV energy range
  previously used for the study of the projected temperature
  profile. These data are obtained using the \texttt{projct*wabs*apec}
  model, fixing the absorbing column density to 2.28$\times$10$^{20}$
  cm$^{-2}$.  Temperature (in keV), abundances (in fraction of solar
  value) and normalizations are left as free parameters. The first
  central 1$''$.5 are excluded from the analysis.  The result of the
  fit gives $\chi^{2}$/dof $=$ 1418.40/1400.  Densities have been
  derived from the \texttt{apec} normalization:
  10$^{-14}$n$_{e}$n$_{H}$V/4$\pi$[D$_{A}$(1+z)]$^{2}$, assuming
  \textit{$n_{p}$ = 0.82n$_{e}$}.  Errors are at 90\% of confidence
  level, while those enclosed in brackets are at 1$\sigma$.}
\end{deluxetable*}

\section{Cool Core Analysis} 
\label{sec:Cool Core Analysis}

\begin{figure}[ht]
\includegraphics[width=0.48\textwidth]{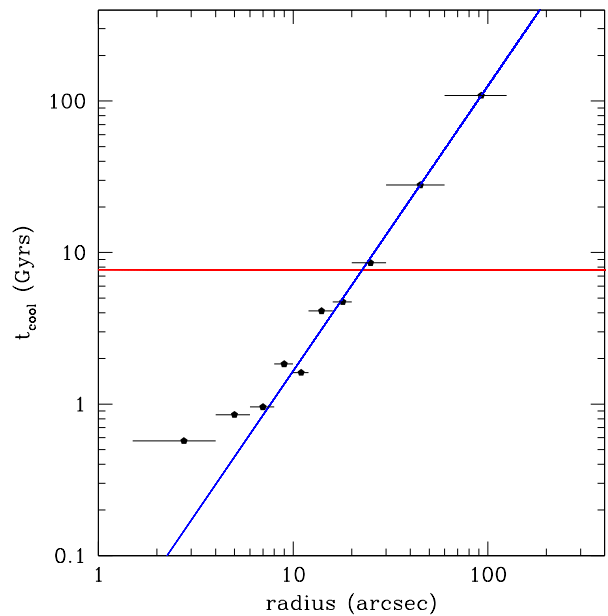}
\caption{Cooling time profile measured from the spectral deprojected
  analysis.  The red line indicates the value of 7.7 $\times$ 10$^{9}$
  yr which is assumed to be the cluster age.  The blue line is the
  powerlaw fit to the cooling time radial profile.  The corresponding
  value of the cooling radius is r$_{cool} \simeq$ 22$''$.7 ($\simeq$109
  kpc).}
			\label{fig:cool_rad}
\end{figure}

The X-ray temperature and surface brightness profiles show the
evidence of the presence of a cool core in RBS 797.  We estimated the
cooling time as the time necessary to the ICM to radiate its enthalpy
per unit of volume:
\begin{equation}\label{eq:cooling time}
t_{cool} \approx \frac{H}{n_{e}n_{H}\Lambda(T)} =
\frac{\gamma}{\gamma - 1}\ \frac{kT}{\mu X_{H}n_{e}\Lambda(T)}
\end{equation}
where $\mu$ $\approx$ 0.61 is the molecular weight for a fully ionized
plasma, $\gamma$ = 5/3 is the adiabatic index, \textit{X$_{H}$}
$\approx$ 0.71 is the hydrogen mass fraction and \textit{$\Lambda$(T)}
is the cooling function.
By adopting the values of the temperature and electron density derived
from the spectral deprojection analysis (see Section
\ref{subsec:Deprojection Analysis}), we derive the radial profile of
the cooling time (Figure \ref{fig:cool_rad}), which shows a power law
behaviour. The fit with a power law gives $t_{cool}$ $\propto$
$r^{1.88}$. The central cooling time, estimated in the first bin
(1$''$.5--4$''$), is $t_{cool}$ = 5.7 $\times$ 10$^{8}$ yr.

It is also possible to estimate the cooling radius, defined as the
radius within which the gas has a cooling time less than the age of
the system. The choice of the method adopted to estimate such an age
is arbitrary.  For consistency with the assumption usually made for
local clusters, we adopt the cluster's age to be 7.7 $\times$ 10$^{9}$
yr (the look back time at the epoch z = 1, at which many clusters
appear to be relaxed and a cooling flow could establish itself).  We
thus estimate r$_{cool} \simeq$ 22$''$.7 (Figure \ref{fig:cool_rad}),
corresponding to $\simeq$ 109 kpc. Our estimate of $r_{cool}$ is
slightly different from that of \cite{Rafferty_2006} and
\cite{Cavagnolo_2011} due to the different definition of $t_{cool}$
used.

The spectrum extracted within the cooling radius (excluding the
central 1$''$.5), was fitted with \texttt{XSPEC} using a deprojected
multi-phase model (\texttt{projct*wabs*(apec+mkcflow)}). The only
difference from the model adopted in the deprojection analysis in
Sect. \ref{subsec:Deprojection Analysis} is the addition of an
isobaric multi-phase component (\texttt{mkcflow}).  The \texttt{apec}
parameters of temperature and abundance were left free to vary, as
well as the absorbing column density which was tied between all the
annuli. For the \texttt{mkcflow} model, the high temperature (highT)
and abundance were tied to the corresponding values for the
\texttt{apec} model, for which the parameters are free. On the other
hand, the low temperature (lowT) was fixed at $\sim$0.1 keV as we are
assuming the physics of a standard cooling flow model.

The best-fit parameters give a mass deposition rate of the cooling
flow, calculated from the \texttt{mkcflow} normalization, of
~$\dot{M}$ $=$ 231$^{+316}_{-227}$ M$_{\odot}$ yr$^{-1}$ (consistent
with \citealp{Rafferty_2006}), a deprojected cooling flow temperature
of kT$_{1''.5-22.7''}$ = 5.77$^{+0.42}_{-0.38}$ keV and a cooling
X-ray luminosity inside the cooling radius within the energy range
0.5--7 keV of L$_{X,~cool}$ = 1.33 $\times$ 10$^{45}$ erg s$^{-1}$.
The high nominal value of the mass deposition rate that we estimate
in the framework of the standard cooling flow model is likely due to the
temperature gradient seen in the profile (left panel of Figure
\ref{fig:rad_met_temp}), since the temperature within the cooling radius is
clearly lower than the values found at outer radii. Nevertheless, for
comparison with the literature which adopted the standard cooling flow model,
we use this value of the mass deposition rate for estimating the cooling flow
power presented in Section \ref{minihalo.sec}.
%%%%%%%%%%%%%%%%%%%%%%%%%%%%%%%%%%%%%%%%%%%%%%%%%%%%%%%%%%%%%%%%%%%%%%%%%%%%%%

\section{Discussion} 
\label{Discussion}

\subsection{Energetics} 
\label{subsec:energetics}

The comparison between the X-ray luminosity of the ICM within the
cooling radius and the cavity power provides us with an estimate of
the balance between the energy losses of the ICM by the X-ray emission
and the cavity heating \citep{Birzan_2004, Rafferty_2006}.  The
bolometric X-ray luminosity inside the cooling region is estimated
from a spectral deprojection analysis as L$_{X,~cool}$ $=$ 1.33
$\times$ 10$^{45}$ erg s$^{-1}$, consistent with the estimate of
\cite{Rafferty_2006}.

The power of the cavities P$_{cav}$ is given by the ratio of the
cavity energy and the cavity age.  The energy required to create a
cavity with pressure $p$ and volume $V$ is the sum of the $pV$ work
done by the jet to displace the X-ray emitting gas while it inflates
the radio lobes, and the internal energy $E$ of the cavity system.
This quantity is the enthalpy defined as:
\begin{equation}
E_{\rm cav} = H = E + pV = \frac{\gamma}{\gamma - 1} pV
\end{equation}
where $\gamma = 4/3$ for relativistic plasma. Pressure and volume can
be estimated directly by X-ray observations through measurements of
the cavity size and of the temperature and density of the surrounding
ICM.  A potential issue is the uncertainty in determinations of the
cavity volumes. The cavity size is usually estimated through a visual
inspection of the X-ray images. This method is therefore dependent on
the quality of the X-ray data, and is also prone to systematic error.
Furthermore, projection affects the determination of the
actual spatial geometry of the cavity system.  The cavity size and
geometry measured by different observers may vary significantly
depending on the approach adopted, leading to differences between
estimates of up to a factor of few in $pV$ \citep[e.g.,][]{Gitti_2010,
  O'Sullivan_2011, Cavagnolo_2011}. Here we estimate the cavity size
from the surface brightness profile of the central region of the
cluster.  From the decrement of the surface brightness observed within
the two X-ray deficient regions containing the lobes (Figure
\ref{fig:SB}), we estimated the extent of the major axis of the
two ellipses. The surface brightness in the directions of the cavities
clearly falls below that of the undisturbed cluster within the
$\sim$2$''$--7$''$ radius.

For the estimate of the ICM pressure ($p \simeq 1.92 n_{e}kT$)
surrounding the cavities, we considered the electron density and
temperature within the bin 4$''$--6$''$ in the cluster profiles (left
and right panels of Figure \ref{fig:rad_met_temp}).  We calculated
that the total enthalpy of the cavity system is E$_{cav}$ = 1.68
$\times$ 10$^{60}$ erg.  For the estimate of the cavity age, we
adopted the sound crossing time, defined as $t_{sound}$ $\approx$
R/$c_{s}$, where R is the distance of the center of the cavity from
the central radio source and c$_{s}$ is the sound speed.  Finally, we
find total cavity power of P$_{cav}$ = 2.49 $\times$ 10$^{45}$ erg
s$^{-1}$. The cavity properties are listed in Table \ref{tab:cavity
  energetics}.

Although we estimate slightly smaller cavity volumes and we adopt a
different assumption for the cavity age (sound crossing time instead
of buoyancy time), we find values of energy and power of the cavities
consistent with those of the \textit{configuration-I} in
\cite{Cavagnolo_2011}.

\begin{deluxetable}{lcc}[b]
\tabletypesize{\scriptsize}
\tablewidth{0pt}
\tablecaption{ Cavity properties
\label{tab:cavity energetics}
 }
\tablehead{
\colhead{} & \colhead{Cavity NE} & \colhead{Cavity SW}}
\\
\startdata
$a$~ ($''$)  & $2.67$ & $2.70$
\\
$b$~ ($''$)  & $2.30$ & $2.29$
\\
$R$~ ($''$)  & $5.37$ & $4.62$
\\
$a$~ (kpc)  & $12.82$ & $12.96$
\\
$b$~ (kpc)  & $11.04$ & $10.99$
\\
$R$~ (kpc)  & $25.78$ & $22.18$
\\
$V \, ({\rm cm}^3)$ & $1.93 \times 10^{68}$ & $1.93 \times 10^{68}$
\\
$p \, {\rm (erg \, cm}^{-3})$ & $1.09 \times 10^{-9}$
& $1.09 \times 10^{-9}$ 
\\
$kT$ (keV) & $4.61$ & $4.61$
\\
$p V$~ (erg)& $2.10 \times 10^{59}$ & $2.10 \times 10^{59}$
\\
$E_{\rm cav} \, (\rm erg )$ & $8.41 \times 10^{59}$ & $8.41 \times
10^{59}$
\\
$t_{\rm sound}$~ (yr) & $2.31 \times 10^{7}$ & $2.02 \times 10^{7}$ 
\\
$P_{\rm cav} \, ({\rm erg \, s}^{-1})$ & $1.15 \times
10^{45}$ & $1.34 \times 10^{45}$
\\[-2.5mm]
\enddata
\tablecomments{Results of the analysis of the cavity system of RBS
  797. We assume the shape of the cavities as prolate ellipsoids. The
  cavity energy is measured as $E_{cav} = 4pV$ and the cavity power is
  estimated as $P_{\rm cav}=E_{cav}/t_{\rm sound}$.}
\end{deluxetable}

%%%%%%%%%%%%%%%%%%%%%%%%%%%%%%%%%%%%%%%%%%%%%%%%%%%%%%%%%%%%%%%%%%%%%%%%%%%%%%

\subsection{Evidence for metal-enriched outflows}
\label{subsec:metal}
\begin{figure}[h]
\includegraphics[width=0.48\textwidth]{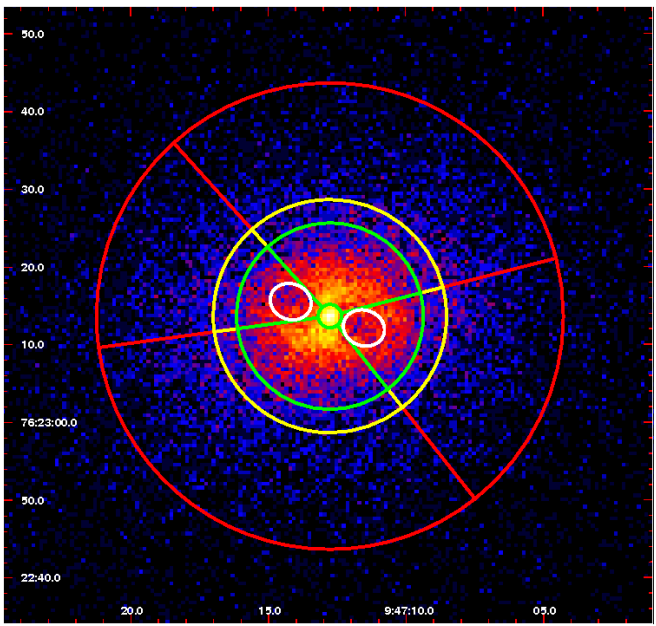}

\caption{Raw image of RBS 797 in the 0.5--7.0 keV energy range showing
  the subdivision in the four wedges used in the study of the iron
  abundance variations.  The white ellipses indicate the cavities.
  The four sectors are tangent to the cavities and divide the cluster
  in two unperturbed regions (N and S) and two cavity regions (NE and
  SW). The angular apertures of each region is: NE$\sim$56 $^\circ$,
  SW$\sim$66$^\circ$, N$\sim$118$^\circ$, S$\sim$120$^\circ$.  The
  measurements of the metallicity were taken in three annuli which
  differs only for the length of the external radius (see Table
  \ref{tab:abundance_variations}).  The green sectors indicate the
  1$''$.5--12$''$ selection while the yellow indicate the
  1$''$.5--15$''$ bin, and the more external red lines represent the
  biggest radius at 1$''$.5--30$''$. }
			\label{fig:tangent_sectors}
\end{figure}
Hydrodynamic simulations \citep[e.g.][]{Bruggen_2002, Roediger_2007,
  Mathews-Brighenti_2008} predicted that in cool core clusters with a
cavity system, the ICM in the core is stirred. Jets from the central
source in clusters of galaxies can contribute both to quench the
deposition of cold gas and to induce motions affecting the metallicity
profile.  Moreover, \cite{Gaspari_2011b, Gaspari_2011a} show new
detailed models of AGN-injected jets which uplift the metal-enriched
gas in the core by feedback cycles. These are split in two phases
during the outburst: initially the gas with higher abundance lies
asymmetrically along the jet axis, followed by a phase of turbulence
and mixing.  \cite{Gaspari_2011b, Gaspari_2011a} estimate that the maximum
distance along the cavity axis at which the enriched gas could be spread is
$\sim 40$ kpc, and that the contrast of the iron abundance in the
cavity directions with that in the rest of the ICM is $\sim$ 0.1--0.2.

In RBS 797 we find differences in the iron abundance distribution,
depending on direction.  We considered four sectors delimited by a
variable outer radius starting from the same 1$"$.5 inner radius (see
Figure \ref{fig:tangent_sectors}).  In our analysis we consider the N
and the S sectors as unperturbed, since they enclose only the regions
of the cluster in which the ICM distribution does not show special
features such as the cavities.  The first analysis is performed in the
annulus from 1$''$.5 to 12$''$ while the second outer radius was set
at 15$''$ and the last at 30$''$.
As for the study of the azimuthally-averaged temperature profile, the
temperature, abundance and normalization were left free to vary, while
redshift and gas column density were fixed to the same values
previously used in Section \ref{subsec:Azimuthally Averaged
  Profiles}. Spectra were fitted with a single plasma model for a
collisionally-ionized diffuse gas.

\begin{deluxetable*}{cccccccc}[t]
\tablewidth{0pt}
\tablecaption{ Abundance variations
\label{tab:abundance_variations}
 }
\tablehead{
\colhead{Sectors} & \colhead{r1 - r2} & \multicolumn{2}{c}{r} &
\colhead{Source} & \colhead{kT} & \colhead{Z} &	{$\chi^{2}$/dof} \\
\colhead{($^\circ$)} & \colhead{($''$)} & \colhead{($''$)} & \colhead{(kpc)}
& \colhead{Counts} & \colhead{(keV)} & \colhead{(Z$_{\odot}$)} & {}
}
\startdata
NE (cav) [$\sim$56$^\circ$]	&	1.5 - 12	&	 6.75	&
32.4 &	2381	&	5.36$_{-0.52~(0.32)}^{+0.72~(0.41)}$	&
0.82$_{-0.33~(0.24)}^{+0.38~(0.22)}$	&	66.3/73 \\[+1mm]
		&	1.5 - 15		&	8.25	&	39.6
& 2996 &	5.77$_{-0.55~(0.38)}^{+0.71~(0.43)}$	&
  0.65$_{-0.26~(0.16)}^{+0.28~(0.17)}$	&	88.1/90\\[+1mm]
		  &	1.5 - 30		&	15.75	&	75.6
&	4886	&	5.91$_{-0.53~(0.33)}^{+0.58~(0.34)}$
&	0.57$_{-0.19~(0.11)}^{+0.20~(0.12)}$	&	123.1/143 \\[+1mm]
		\hline\\[-2mm]
		N [$\sim$118$^\circ$]	&	1.5 - 12	&
6.75	&	32.4	&	6154	&
4.98$_{-0.33~(0.20)}^{+0.33~(0.21)}$	&
0.37$_{-0.13~(0.08)}^{+0.13~(0.08)}$	& 134.0/164\\[+1mm]
&	1.5 - 15	&	8.25	&	39.6	&	7307	&
5.26$_{-0.32~(0.19)}^{+0.39~(0.11)}$	&
0.37$_{-0.12~(0.07)}^{+0.12~(0.07)}$	&	162.3/180 \\[+1mm]
&	1.5 - 30	&	15.75	&	75.6	&
10622 &	5.84$_{-0.38~(0.10)}^{+0.39~(0.10)}$	&
0.35$_{-0.10~(0.06)}^{+0.10~(0.06)}$	&	186.3/221 \\[+1mm]
		\hline\\[-2mm]
		SW (cav) [$\sim$66$^\circ$]	&	1.5 - 12	&
6.75	&	32.4	&	3363	&
5.25$_{-0.45~(0.27)}^{+0.62~(0.34)}$	&
0.61$_{-0.22~(0.14)}^{+0.24~(0.14)}$ & 77.1/102\\[+1mm]
&	1.5 - 15	&	8.25	&	39.6	&	4121	&
5.27$_{-0.40~(0.24)}^{+0.56~(0.31)}$	&
0.66$_{-0.20~(0.13)}^{+0.22~(0.13)}$ 	&	104.9/122\\[+1mm]
		  &	1.5 - 30		&	15.75	&	75.6
&	6319	&	5.96$_{-0.47~(0.29)}^{+0.49~(0.28)}$	&
0.47$_{-0.15~(0.09)}^{+0.15~(0.09)}$	&	126.2/171\\[+1mm]
		\hline\\[-2mm]
		S [$\sim$120$^\circ$]	&	1.5 - 12	&
6.75	&	32.4	&	6661	&
5.21$_{-0.32~(0.20)}^{+0.40~(0.20)}$ &
0.50$_{-0.14~(0.08)}^{+0.15~(0.09)}$	& 199.1/172
\\[+1mm]
&	1.5 - 15	&
8.25
&	39.6	&	7873	&
5.31$_{-0.30~(0.18)}^{+0.42~(0.22)}$	&
0.48$_{-0.12~(0.07)}^{+0.13~(0.08)}$	&	219.9/191 \\[+1mm]
		  &	1.5 - 30		&	15.75	&	75.6
&	11349	&	6.04$_{-0.36~(0.22)}^{+0.37~(0.22)}$	&
0.38$_{-0.10~(0.06)}^{+0.10~(0.06)}$	&	253.1/235\\[+1mm]
		\hline
		\hline\\[-2mm]
		NE + SW	&	1.5 - 12	&	6.75	&	32.4
&	5743	&	5.31$_{-0.34~(0.21)}^{+0.47~(0.26)}$
&	0.68$_{-0.19~(0.12)}^{+0.20~(0.12)}$	& 144.0/177\\[+1mm]
&	1.5 - 15	&	8.25	&	39.6
&	7286	&	5.60$_{-0.36~(0.24)}^{+0.44~(0.27)}$
&	0.64$_{-0.16~(0.10)}^{+0.17~(0.10)}$	&	219.0/223 \\[+1mm]
		  &	1.5 - 30		&	15.75	&	75.6
&	11487	&	6.03$_{-0.35~(0.22)}^{+0.36~(0.22)}$
&	0.51$_{-0.12~(0.07)}^{+0.12~(0.07)}$	&	290.3/331 \\[+1mm]
		\hline\\[-2mm]
		N + S	&	1.5 - 12	&	6.75	&	32.4
&	12812	&	5.10$_{-0.23~(0.14)}^{+0.24~(0.14)}$	&
0.43$_{-0.09~(0.06)}^{+0.10~(0.06)}$	& 335.4/338 \\[+1mm]
&	1.5 - 15	&	8.25	&	39.6
&	15628	&	5.32$_{-0.21~(0.13)}^{+0.28~(0.14)}$
&	0.43$_{-0.09~(0.05)}^{+0.09~(0.05)}$	&	470.3/392\\[+1mm]
		  &	1.5 - 30		&	15.75	&	75.6
&	22541	&	6.00$_{-0.26~(0.16)}^{+0.26~(0.16)}$	&
0.37$_{-0.07~(0.04)}^{+0.07~(0.04)}$	&	570.9/483 \\[-2.5mm]
\enddata
\tablecomments{\textit{Upper section:} results of the spectral
  analysis in the 0.5--7.0 keV energy range of the four sectors
  tangent to the cavities (Figure \ref{fig:tangent_sectors}).
  \textit{Lower section:} results obtained by extracting the combined
  spectra of the two sectors containing the cavities (NE and SW), and
  the two undisturbed sectors (N and S), respectively.  The first
  column lists the delimiting radii for each annulus, while the second
  and the third columns list the central value for each bin.  Errors
  are at 90\% of confidence level, while those enclosed in brackets
  are at 1$\sigma$.}
\end{deluxetable*}

\begin{figure}[htb]
\includegraphics[width=0.38\textwidth]{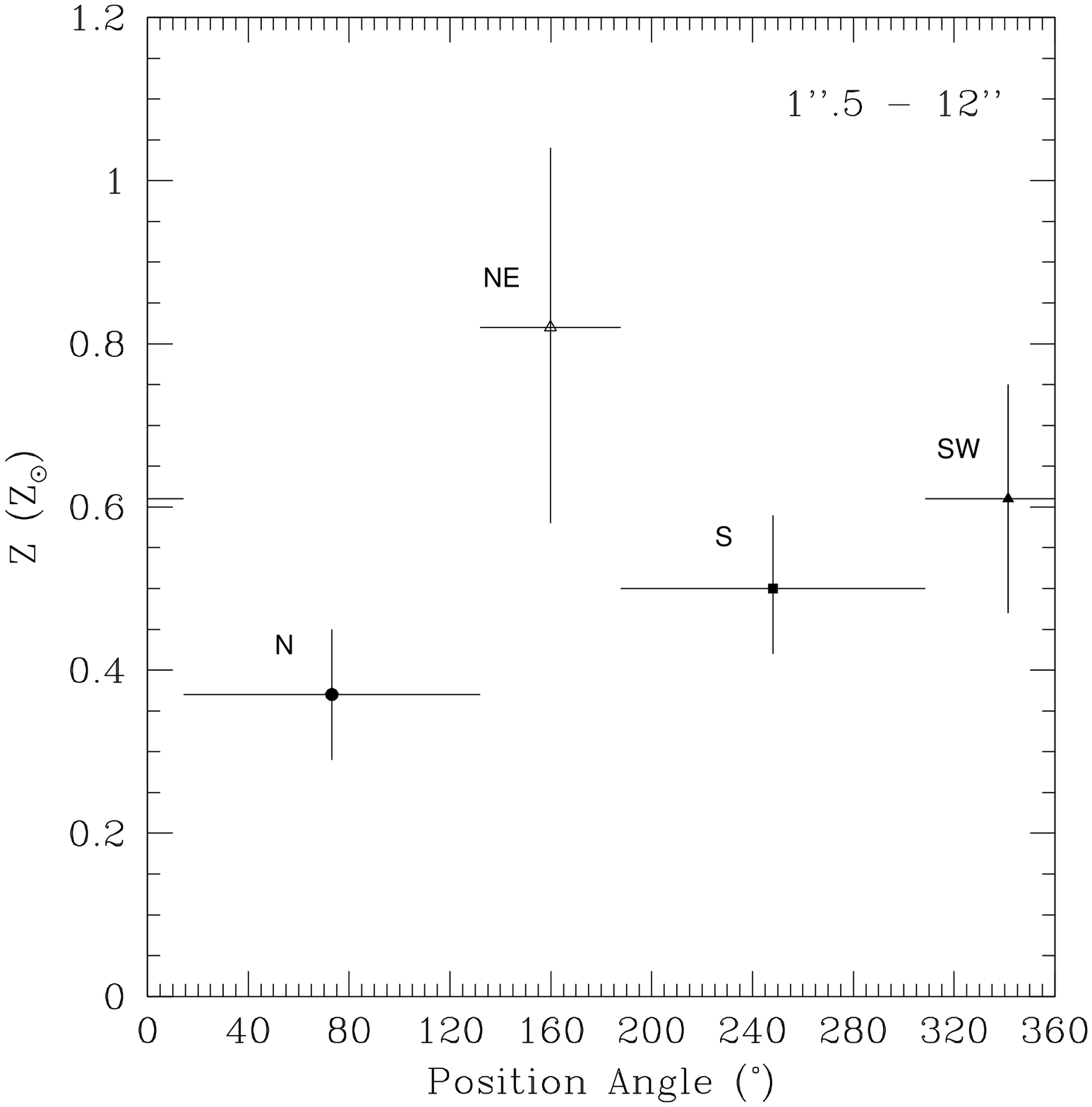}
\includegraphics[width=0.38\textwidth]{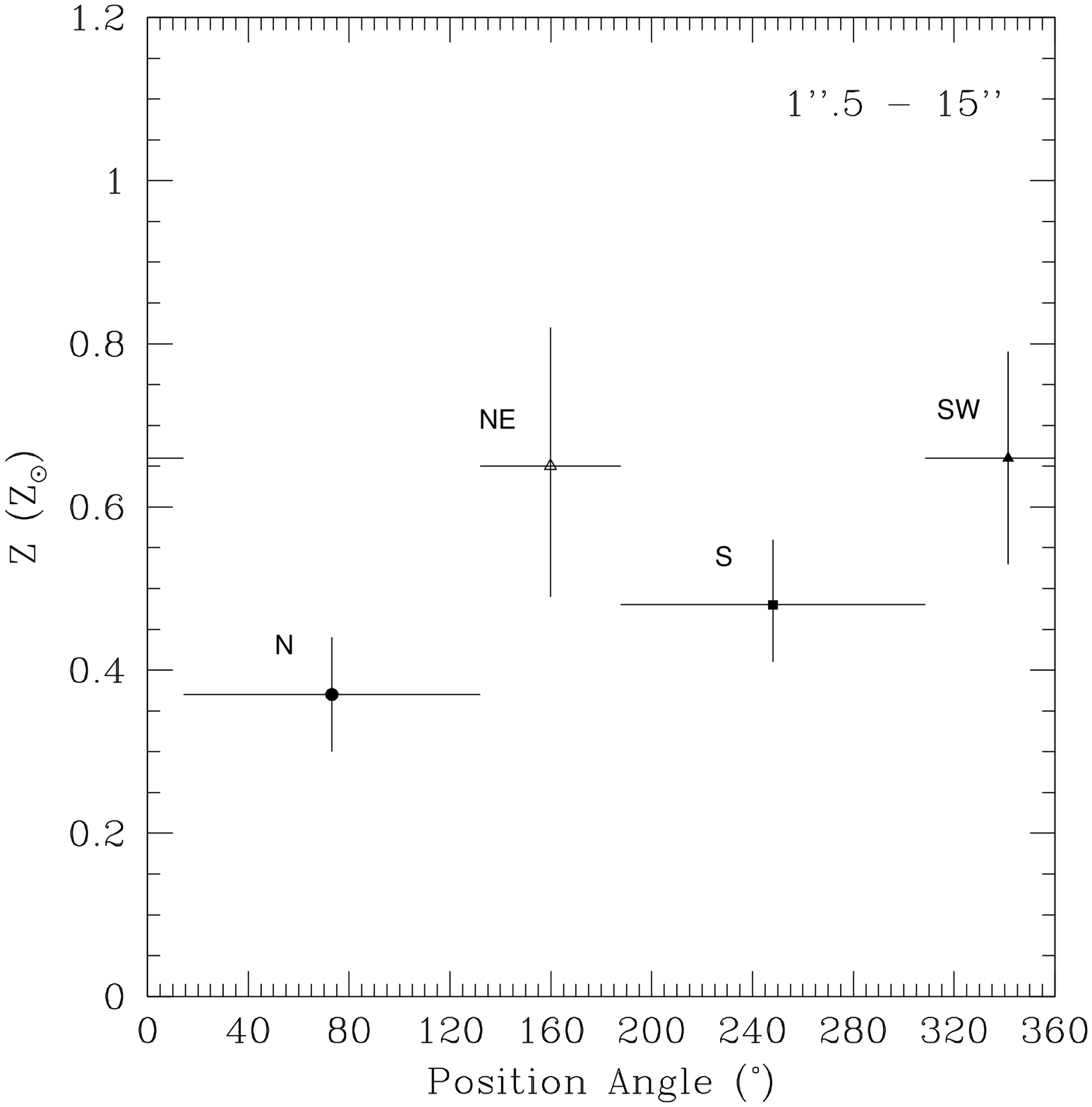}
\includegraphics[width=0.38\textwidth]{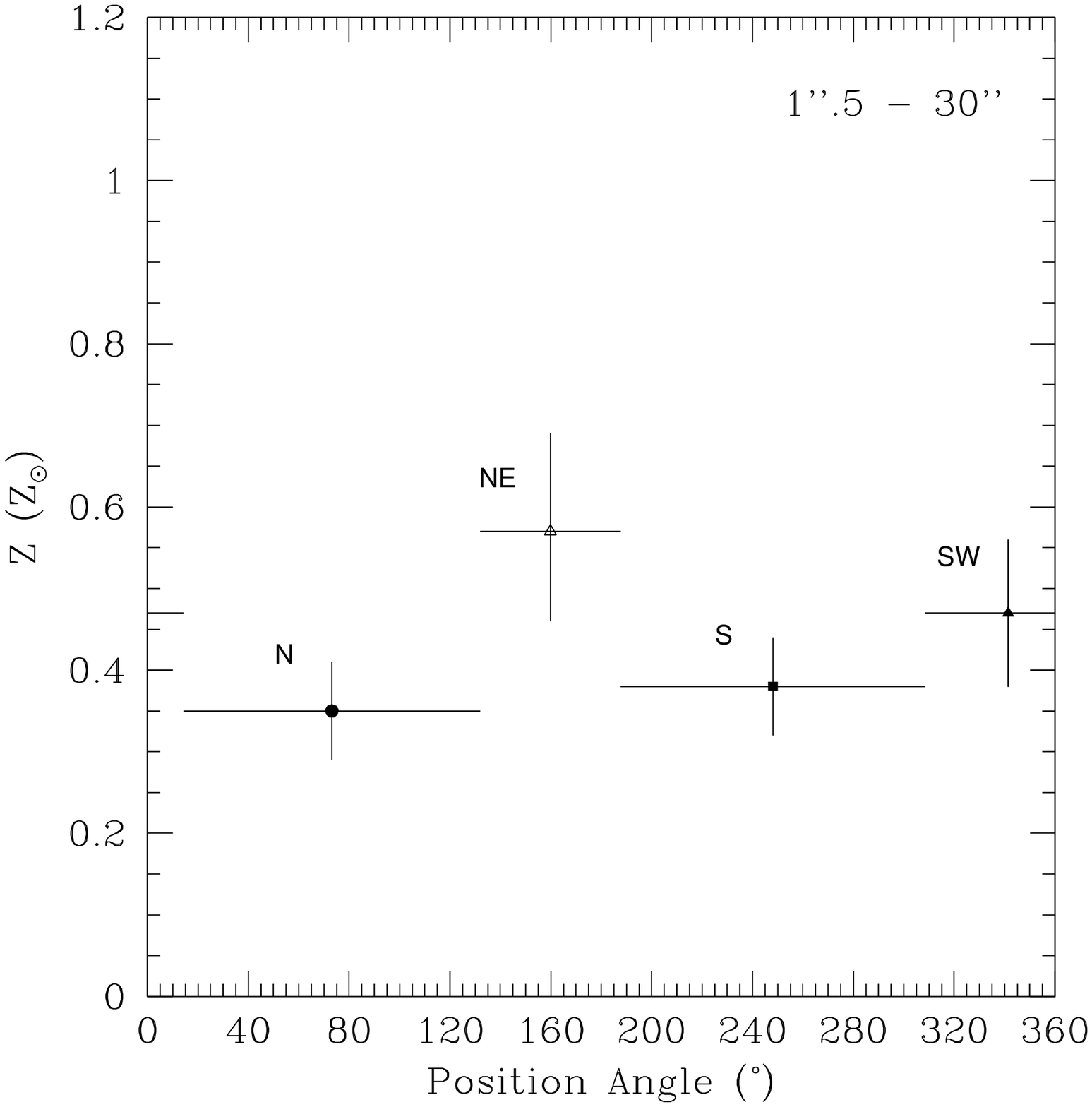}
\caption{Metallicity values (from Table
  \ref{tab:abundance_variations}) as a function of the mid-aperture
  angle of the sector in three different annuli: 1$''$.5--12$''$ (left
  panel), 1$''$.5--15$''$ (central panel), 1$''$.5--30$''$ (right
  panel).  The 0$^\circ$ angle is defined along the W direction and
  the aperture angle increases counterclockwise. Here the four wedges
  are taken singularly: the full triangle indicates the value in the
  SW cavity, the circle that in the N region, the empty triangle that
  in the NE cavity and square that in the S region. The regions
  considered are shown in Figure \ref{fig:tangent_sectors}. Errorbars
  are at 1$\sigma$.}
			\label{fig:abundance_variations}
\end{figure}

The results in Table \ref{tab:abundance_variations} and Figure
\ref{fig:abundance_variations} indicate higher metallicity along the
NE and SW cavity directions compared to the unperturbed N sector in
all the three annuli (at 1$\sigma$ error). In the biggest annulus
(1$''$.5--30$''$), the NE sector shows higher abundance compared to
both the unperturbed N and S regions.
Combining spectra from the two cavity sectors in a joint fit and
comparing to the results of fitting the two unperturbed sectors
supports the relationship between metallicity and direction (see lower
section of Table \ref{tab:abundance_variations} for details).

Likely, the AGN outburst may have lifted the cooler, metal-rich gas
from the central region to larger radii along the directions of the
inflating cavities.  We found that, taking into account three
different outer radii, the cavity sectors have higher abundances than
the unperturbed regions, at 1$\sigma$ confidence. So, this behaviour
persists also at larger radii and not only nearby the cavity system.
This effect has been found in other systems (e.g. M87,
\citealp{Simionescu_2008}; Hydra A, \citealp{Kirkpatrick_2009},
\citealp{Simionescu_2009b}; NGC 5813, \citealp{Randall_2011}).  In particular,
recent work by \citealp{Kirkpatrick_2011} shows a correlation between jets and
metallicity.

%%%%%%%%%%%%%%%%%%%%%%%%%%%%%%%%%%%%%%%%%%%%%%%%%%%%%%%%%%%%%%%%%%%%%%%%%%%%%%

\subsection{Evidence for hot cavities} 
\label{subsec:Rims}

\begin{figure}[t]
\includegraphics[width=0.48\textwidth]{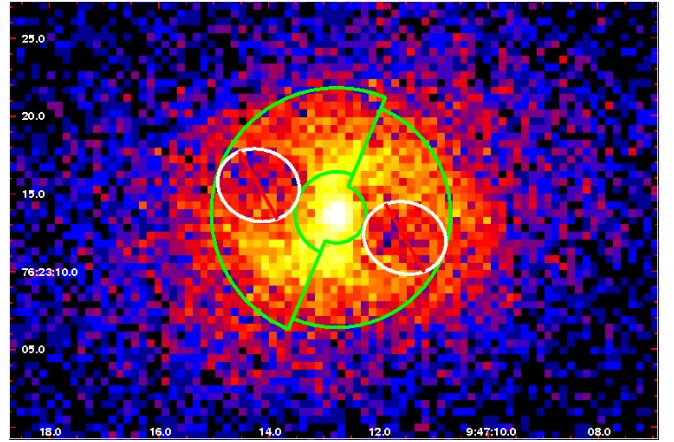}
\caption{Regions considered to estimate the projected temperature in
  the gas (green) surrounding the cavity system, overlayed on the
  0.5--7.0 keV raw image of the core of RBS 797. The dimensions of the
  cavities (white) are listed in Table \ref{tab:cavity energetics}.}
		\label{fig:rims_methods}
\end{figure}

In Section \ref{subsec:metal} we described the analysis of the
azimuthal variations of the ICM properties. Here we want to perform a
more detailed analysis of the projected temperature in the cavities
and in the surrounding gas. We extracted the spectra separately in
both the cavities, and in each sector surrounding a single cavity
(clearly excluding the cavities). Spectra were then fitted with the
model \texttt{wabs*apec} in \texttt{XSPEC}, with the temperature and
abundance as free parameters and column density set at the Galactic
value. Considering the regions of Figure \ref{fig:rims_methods}, in
the cavities we found projected temperatures (1$\sigma$ errors) of
kT$_{cavNE}$ = 5.86$^{+0.91}_{-0.79}$ keV for the NE cavity and
kT$_{cavSW}$ = 5.71$^{+0.92}_{-0.72}$ keV for the SW cavity.  These
have been compared with the kT$_{gasNE}$ = 4.62$^{+0.24}_{-0.23}$ keV
and kT$_{gasSW}$ = 5.08$^{+0.25}_{-0.24}$ keV, respectively the values
for the gas surrounding the NE and the SW cavities.  Since these
results do not provide definite constraints due to the low statistics
of the selected regions, we tried to combine together the two cavities
and compare them to the combined spectra of the external gas. The
results of the joint fits give an indication for hotter cavities:
kT$_{gas}$ = 4.84$^{+0.17}_{-0.17}$ keV and in the cavities kT$_{cav}$
= 5.27$^{+0.59}_{-0.40}$ keV.

We performed an additional test by means of a \textit{``softness ratio''}
(SR), which is defined as $SR = (S-H)/(S+H)$, where $S$ and $H$ are
the count rates in the soft and in the hard band respectively.  After
extracting the surface brightness profile in the cavity system and in
the external gas, both in the X-ray soft (0.5--2.0 keV) and hard
(2.0--7.0 keV) energy band, we compared their softness ratios.
The SR method can be used as an indicator of the temperature, since
higher SR values would imply the prevalence of the soft emission, thus
lower projected temperatures. We found that the regions surrounding
the cavities have a softness ratio slightly higher than that of the
cavities: $SR_{cav} = 0.5468 \pm 0.0216$, $SR_{gas} = 0.5747 \pm
0.0093$. This means that the relative quantity of soft photons is
higher in the outer gas, indicating that the cavities are hotter than
the surrounding ICM.

\begin{figure}[t]
\includegraphics[width=0.48\textwidth]{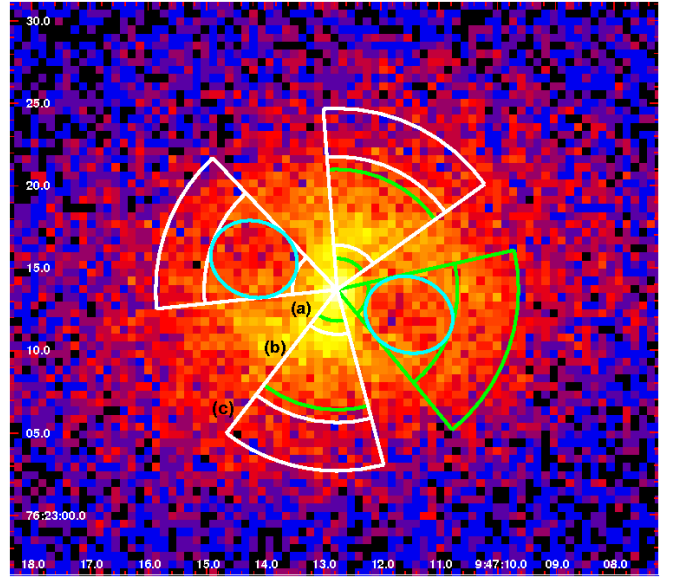}
\caption{Regions adopted to estimate the softness ratio.  Three bins
  are selected in order to calculate the softness ratio in three
  regions: \textit{(a)} from the center of the cluster to the inner
  edge of the selected cavity (cyan ellipses); \textit{(b)} encloses
  the whole cavity; \textit{(c)} from the outer edge of the cavity to
  the end of the rims.  The radii of the three bins are different for
  the two cavities, while the aperture angles are the same in both
  cases.  The bins for the NE cavity (white regions) are: \textit{(a)}
  0$''$--2$''$.7; \textit{(b)} 2$''$.7--8$''$.1; \textit{(c)}
  8$''$.1--11$''$.  The bins used for the SW cavity (green
  regions) are: \textit{(a)} 0$''$--1$''$.9; \textit{(b)}
  1$''$.9--7$''$.3; \textit{(c)} 7$''$.3--11$''$.}
			\label{fig:3bin-hr}
\end{figure}

\begin{figure*}[tb]
\includegraphics[width=0.48\textwidth]
{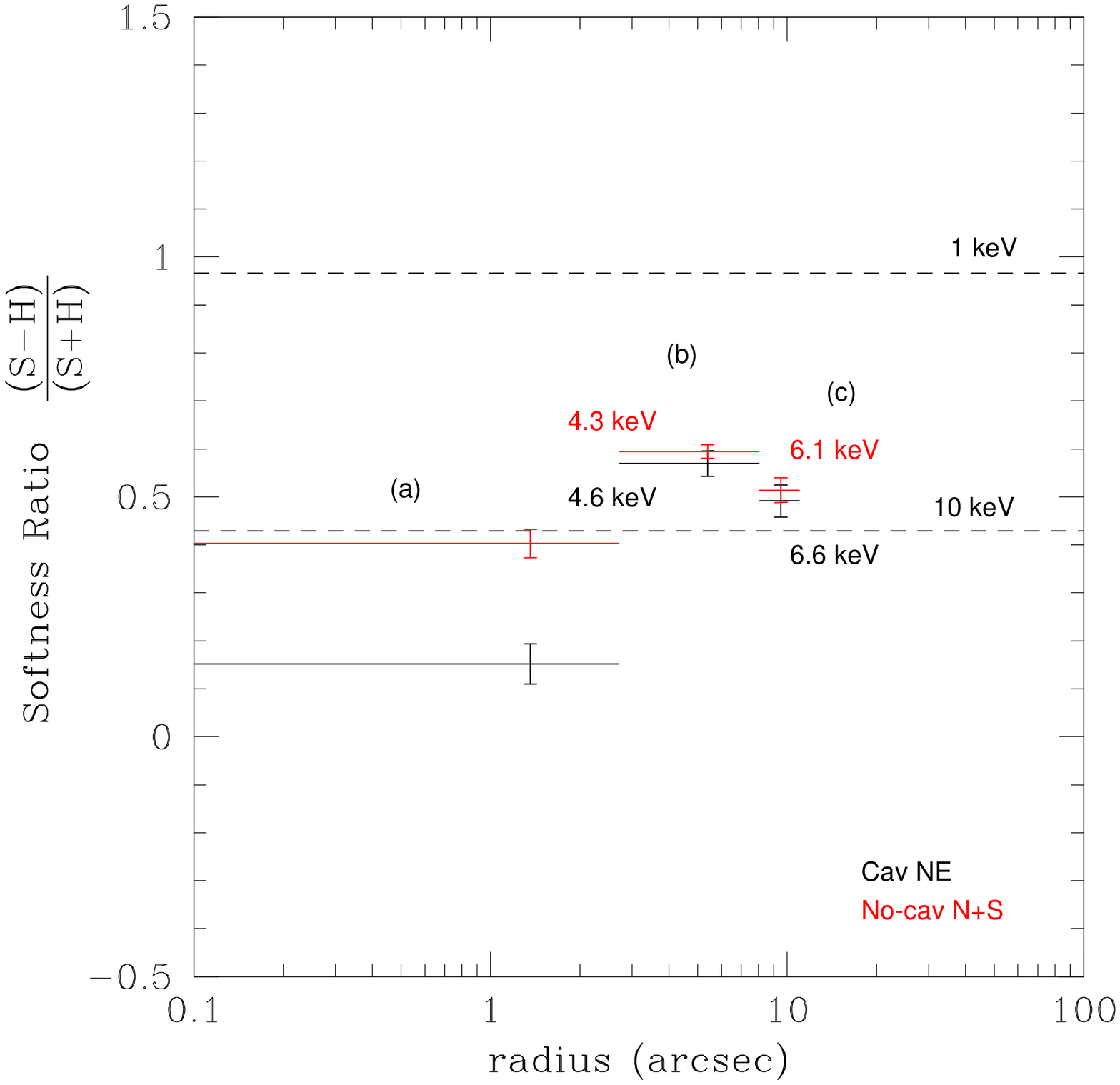}
\includegraphics[width=0.48\textwidth]
{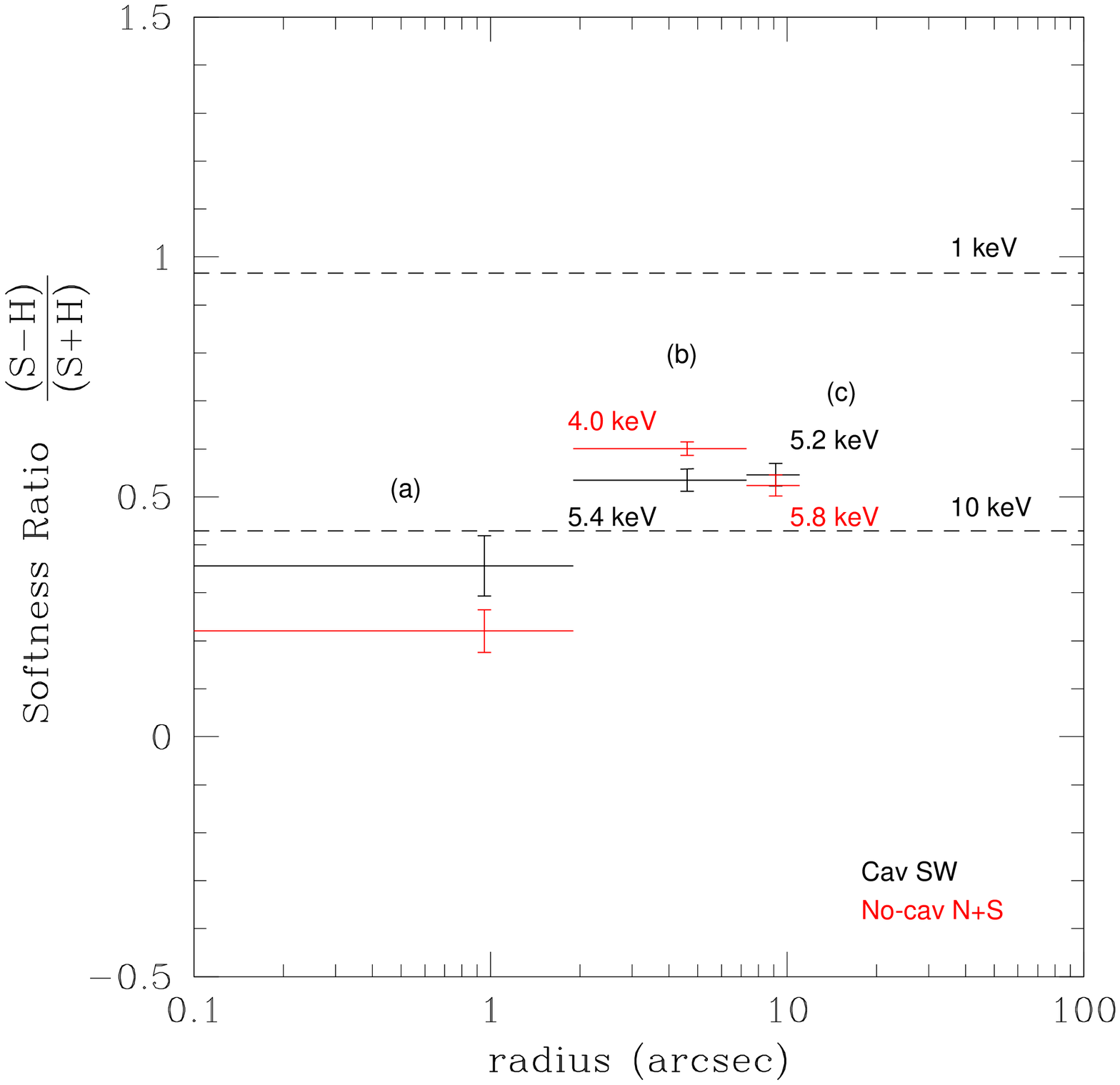}
\caption{Softness ratio as an indicator of the temperature. We have
  considered three bins, different in each cavity direction. The
  wedges corresponding to the \textit{(a)}, \textit{(b)}, \textit{(c)}
  bins are shown in Figure \ref{fig:3bin-hr}.  The dimension of the
  annuli of the unperturbed regions are taken equal to those of the
  region with which are compared, while their angular aperture does
  not vary. The dashed lines give useful hints about the corresponding
  temperatures.  Uncertainties are taken at 1$\sigma$ confidence.}
			\label{fig:2_har_rat}
\end{figure*}

\begin{deluxetable}{cccc}[t]
\tablewidth{0pt}
\tablecaption{ Softness Ratio
\label{tab:3bin-val}
 }
\tablehead{
\colhead{} & \colhead{Bin} & \multicolumn{2}{c}{$\frac{S-H}{S+H}$}\\[+1mm]
\colhead{} & \colhead{} & \colhead{Cavity NE} & \colhead{Cavity SW}
 }
\startdata
	 &	(a)	&	0.1522 $\pm$ 0.0420	&	0.3561 $\pm$
0.0628\\
	Cavities	&	(b)	&	0.5695 $\pm$ 0.0266	&
0.5349 $\pm$ 0.0233\\
	&	(c)	&	0.4915 $\pm$ 0.0335	&	0.5461
$\pm$ 0.0232\\ \hline \\[-2.5mm]
	&	(a)	&	0.4028 $\pm$ 0.0297	&	0.2203 $\pm$
0.0444\\
	No-Cavities	&	(b)	&	0.5948 $\pm$ 0.0141 &
0.6007 $\pm$ 0.0138\\
	&	(c)	&	0.5139 $\pm$ 0.0255	&	0.5242 $\pm$
0.0219\\[-3mm]
\enddata
\tablecomments{Values of the softness ratio calculated in each
annulus. The considered bins (values \textit{(a)}, \textit{(b)}, \textit{(c)}) 
are different for the two cavities and are better visualized in Figure
\ref{fig:3bin-hr}.}
\end{deluxetable}

Finally, we made a further attempt to compare the projected
temperature within the cavities with that of the gas in the other
directions.  We considered four sectors, two along the cavity
directions and two in the perpendicular unperturbed regions, excluding
the cavity rims from the selection.  Basing on the unsharp masked
image (right panel of Figure \ref{fig:raw+unsharp}), we divided
the NE and the SW sectors in three annuli tangent to the inner and
outer edge of the cavities (Figure \ref{fig:3bin-hr}).  The first bin
goes from the centre of the cluster to the beginning of the cavity
(0$''$--2$''$.7 for the NE and 0$''$--1$''$.9 for the SW directions,
\textit{(a)} sectors in Figure \ref{fig:3bin-hr}), the second embodies
the whole cavities (2$''$.7--8$''$.1 for NE and 1$''$.9--7$''$.3 for
SW, \textit{(b)} sectors in Figure \ref{fig:3bin-hr}) and the last
includes the rims (8$''$.1--11$''$ for NE and 7$''$.3--11$''$ for SW,
\textit{(c)} sectors in Figure \ref{fig:3bin-hr}).  Softness ratios
for these regions are listed in Table \ref{tab:3bin-val}.  In Figure
\ref{fig:2_har_rat}, we compare the SR of the unperturbed sectors with
the values of the NE cavity sector (left panel) and with the SW cavity
region (right panel). Looking at the second bin in both the plots of
Figure \ref{fig:2_har_rat}, we find further support for the indication
that the gas in the cavities of RBS 797 is hotter than the surrounding
ICM, as previously found in other cluster of galaxies as 2A 0335+096
\citep{Mazzotta_2003}, MS 0735+7421 \citep{Gitti_2007} and M87
\citep{Million_2010}.

Despite our detailed study of the cavities and surroundings
  temperatures, what we found here may only be a projection
  effect. The hotter gas along our line of sight through the cavities
  could have a greater impact on the spectra of the cavities than on
  those of the brighter gas surrounding them (the rims), making the
  cavities to appear hotter. Furthermore, the cavities are likely to
  be regions from which the (cooler) X-ray emitting gas has been
  pushed out, so a larger fraction of the emission along our line of
  sight may be coming from the projected hotter gas at larger radii.
  We performed a further test in order to investigate whether the
  X-ray emission of the cavities along our line of sight is due
  exclusively to the outer gas emission, or not.  From the surface
  brightness and density profiles (Figure \ref{fig:SB} and right panel
  of Figure \ref{fig:rad_met_temp}, respectively), we found an average
  density of $\sim$0.075 cm$^{-3}$ and a counts estimate of $\sim$5.4
  cts s$^{-1}$ arcmin$^{-2}$ in the annular region containing the
  cavities.  Therefore, at the ambient density and temperature, we
  expect to have $\sim$119 cts s$^{-1}$ with ACIS-S3 chip in the
  0.5--7.0 keV energy band, with the \texttt{apec} normalization
  (K$_{apec}$) set to unity.  Considering a spherical volume
  approximately equal to the volume of the cavities and solving the
  equation for the \texttt{apec} normalization, we estimated
  K$_{apec}$ $\simeq$ 4.78 $\times$ 10$^{-5}$.  From this spherical
  volume of unperturbed gas located at the cavity position we would
  thus measure an emission of $\sim$ 1.04 cts s$^{-1}$ arcmin$^{-2}$
  that, with the exposure time of the ACIS-S3 observation
  (t$_{exp,S3}$ $\simeq$ 36 ks), is equivalent to $\sim$206 counts.
  We considered several regions outside the cavity directions having
  the same cavity area and located at the same distance from the
  center, and we estimated the average number of counts from the
  0.5--7.0 keV image. We found that the difference between the average
  number of counts within these regions ($\sim$1140) and the average
  number of counts inside the cavities ($\sim$766) exceeds the number
  of counts originating from the estimated contribution of the
  unperturbed gas inside the cavities. As a zero-order test, this
  indicates that there is no emission related to the cavities, so that
  they can be considered devoid of X-ray emitting gas. Unfortunately,
  the poor statistics did not allow us to test this hypothesis with
  some more complicated spectral models, such as deprojection
  analysis.

%%%%%%%%%%%%%%%%%%%%%%%%%%%%%%%%%%%%%%%%%%%%%%%%%%%%%%%%%%%%%%%%%%%%%%%%%%%%%%

\subsection{Large-scale radio emission}
\label{minihalo.sec}
\begin{figure}[b]
\vspace{0cm}
\includegraphics[width=8.8cm]{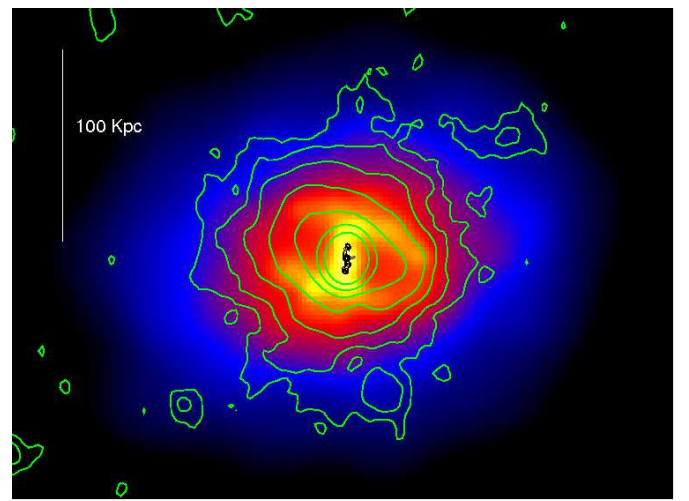}
\vspace{0cm}
\caption{1.4 GHz VLA radio contours at a resolution of $3''$
  (green, see Figure \ref{vla-new.fig}, right panel) and 4.8 GHz VLA
  radio contours at a resolution of $0''.4$ \citep[black, see Figure 2
  of][]{Gitti_VLA} overlaid onto the smoothed \textit{Chandra} X-ray
  image of RBS 797. Smoothing of the \textit{Chandra} ACIS-S3 was
  obtained adaptively in the 0.5--7.0 keV energy range. It was
  obtained by the task \texttt{csmooth} setting the signal-to-noise
  ratio (S/N) at value 3.}
\label{fig:chandra_radio}
\end{figure}
The radio properties of RBS 797 presented in Section
\ref{results-radio.sec} are consistent with a scenario of three
distinct outbursts, with the jet axis precessing $\sim$90$^{\rm
  \circ}$ between outbursts, with a period of $\sim$10$^7$ yr. The
X-ray observations show features that are consistent with this
picture: the northeast-southwest inner cavities and the elliptical
inner edge revealed by the {\it Chandra} images (Figure
\ref{fig:raw+unsharp}) could be associated with the intermediate
outburst; the outer edge shown by {\it Chandra} (Figure
\ref{fig:raw+unsharp}) could be associated with the oldest
outburst.  Alternatively, the observed change in the orientation of
the radio axis, most evident between the inner 4.8 GHz jets and the
1.4 GHz emission filling the cavities (black and green contours in
Figure \ref{fig:chandra_radio}, right panel, see also Fig. 4 of
\citealt{Gitti_VLA}), may be due to jet deflection. Jet deflection is
caused either by ICM pressure gradients, or by dense regions of cold
gas that may have been ionized by massive stars born in a starburst
triggered by the cooling flow \citep[e.g.,][]{Salome&Combes_2003}. The
existing data do not allow us to discriminate between these two
possibilities.

On the other hand, the fact that the total size of the large-scale
radio emission is comparable to that of the cooling region suggests a
different interpretation,
which would be consistent with either jet precession or jet
deflection: that RBS 797 hosts a diffuse radio mini-halo.  Mini-halos
are diffuse radio sources, extended on a smaller scale (up to $\sim
500$ kpc), surrounding a dominant radio galaxy at the cluster center.
They are generally only observed in cool core clusters and it has been
found that these sources are not connected to ongoing cluster major
merger activity.  Their radio emission is indicative of the presence
of diffuse relativistic particles and magnetic fields in the ICM,
since these sources do not appear as extended lobes maintained by an
AGN, as in classical radio galaxies \citep*{Giovannini_2002}.  In
particular, due to the fact that the radiative lifetime of
radio-emitting electrons ($\sim 10^8$ yr) is much shorter than any
reasonable transport time over the cluster scale, the relativistic
electrons responsible for the extended radio emission from mini-halos
need to be continuously re-energized by various mechanisms associated
with turbulence in the ICM (reaccelerated {\it primary} electrons), or
freshly injected on a cluster-wide scale (e.g. as a result of the
decay of charged pions produced in hadronic collisions, {\it
  secondary} electrons).

\begin{figure}[t]
     
\includegraphics[width=0.48\textwidth]{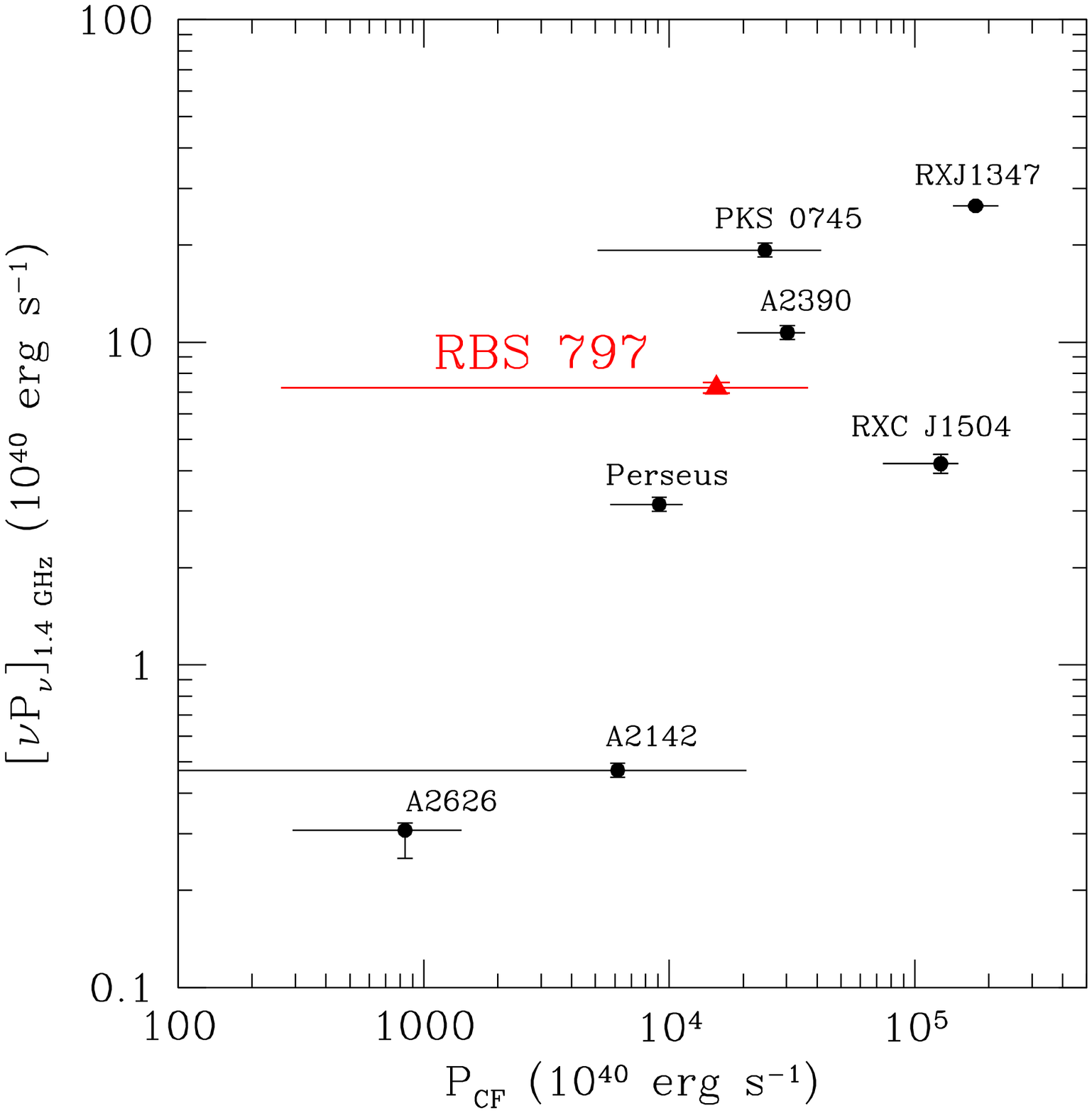}

\caption{Cooling flow power, calculated as P$_{CF}$ =
  $\dot{M}$kT/$\mu$m$_{p}$, versus the integrated radio power $\nu
  P_{\nu}$ at 1.4 GHz. The other objects plotted with RBS 797 (red
  triangle) are galaxy clusters whose diffuse radio emission size is
  comparable to the cooling radius, all taken from the literature
  \citep{Gitti&Schindler_2004, Gitti_2007, Boeringer_2005,
    Giacintucci_2011a}. When necessary, the values have been rescaled
  to the cosmology adopted in this paper.}
			\label{fig:pcf_radpow}
\end{figure}

\cite{Gitti_2002} developed a theoretical model which accounts for the
origin of radio mini-halos as related to electron re-acceleration by
MHD turbulence in cooling flows.  In this model, the necessary
energetics to power radio mini-halos is supplied by cooling flows
themselves, through the compressional work done on the ICM and the
frozen-in magnetic field. This supports a direct connection between
cooling flows and radio mini-halos.  Although secondary electron
models have been proposed to explain the presence of their persistent,
diffuse radio emission on large-scale in the ICM
\citep[e.g.,][]{Pfrommer-Ensslin_2004, Keshet-Loeb_2010}, the observed
trend between the radio power of mini-halos and the maximum power of
cooling flows (Figure \ref{fig:pcf_radpow}) has given support to a
primary origin of the relativistic electrons radiating in radio
mini-halos, favored also by the successful, detailed application of
the model of \cite{Gitti_2002} to two cool core clusters
\citep[Perseus and A 2626,][]{Gitti_2004} and by recent statistical
studies \citep{Cassano_2008}.  However, the origin of the turbulence
necessary to trigger the electron reacceleration is still debated.
The signatures of minor dynamical activity have recently been detected
in some mini-halo clusters, thus suggesting that additional or
alternative turbulent energy for the reacceleration may be provided by
minor mergers \citep{Gitti_2007b, Cassano_2008} and related gas
sloshing mechanism in cool core clusters
\citep{Mazzotta-Giacintucci_2008, ZuHone_2011}.  Given the prevalence
of mini-halos in clusters with X-ray cavities, another attractive
possibility is that the turbulent energy is provided by a small
fraction of the energy released by the lobes rising from the central
AGN \citep[as suggested by][]{Cassano_2008}.

We can test qualitatively the consistency of the observational X-ray
and radio properties of RBS 797 with the trend between the radio power
on the mini-halos and the maximum power of the cooling flows expected
by the reacceleration model. This has been observed in a first sample
of mini-halos selected by Gitti et al. (2004), who found that the
strongest radio mini-halos are associated with the most powerful
cooling flows.  In Figure \ref{fig:pcf_radpow} we show with a red
triangle the radio power at 1.4 GHz of the diffuse emission (in terms
of integrated radio luminosity $\nu P_{\nu}$) versus the maximum power
of the cooling flow $P_{\rm CF} = \dot{M} k T/\mu m_{\rm p}$ (here
$\dot{M}$ is the mass accretion rate, $k$ the Boltzmann constant, $T$
the global ICM temperature, $\mu \approx 0.61$ the molecular weight,
and $m_p$ the proton mass) in RBS 797, overlayed onto the values
measured for the mini-halo clusters known so far. The monochromatic
radio power at the frequency $\nu$ is calculated as
\begin{equation}
P_{\nu}=4 \pi \, D_{\rm L}^2 \, S_{\nu} \, (1+z)^{-(\alpha +1)}
\label{radio_power.eq}
\end{equation}
where $D_{\rm L}$ is the luminosity distance, $S_{\nu}$ is the flux
density at the frequency $\nu$, and $(1+z)^{-(\alpha +1)}$ is the
\textit{K-correction} term (Petrosian \& Dickey 1973), which in the
case of RBS 797 is negligible \citep[$\alpha \sim -1$,][]{Gitti_VLA}.
The radio spectral index $\alpha$ is defined so that $S_{\nu} \propto
\nu^{-\alpha}$.  From the flux density measured in Section
\ref{results-radio.sec}, we estimate a radio mini-halo power of
$P_{\rm 1.4 GHz}= (5.2 \pm 0.2) \times 10^{24}$ W Hz$^{-1}$. From the
global cluster temperature derived in Section \ref{sec:Spectral
  Analysis} and the mass accretion rate $\dot{M}$ derived in Section
\ref{sec:Cool Core Analysis}, which are measured consistently
  with the methods adopted for the other objects in Figure
\ref{fig:pcf_radpow} \footnote{in particular, the mass accretion
    rate is estimated in the framework of the standard cooling flow
    model \citep[see][]{Gitti_2004, Gitti_2007b}}, we estimate
$P_{\rm CF} = 1.55^{+2.12}_{-1.53} \times 10^{44}$ erg s$^{-1}$
(errors at 90\% confidence level). Despite its large errorbar, we can
say that RBS 797 agrees with the observed trend, thus indicating that
its diffuse radio emission can be classified as a radio mini-halo.

On the other hand, we stress that the classification of a radio source
as a mini-halo is not trivial: their detection is complicated by the
fact that the diffuse, low surface brightness emission needs to be
separated from the strong radio emission of the central radio
galaxy. Furthermore, the criteria adopted to define mini-halos are
somewhat arbitrary (e.g., total size, morphology, presence of cool
core) and some identifications are still controversial.

\section{Summary and Conclusions}

In this paper we have shown the results of our analysis of new
\textit{Chandra} and \textit{VLA} observations of the
galaxy cluster RBS 797. Both X-ray and radio data reveal a strong
interplay between the central radio source and the ICM, suggesting
that the cool core cluster RBS 797 is a good candidate to investigate
the AGN feedback process.  Summarizing our main findings:

\begin{enumerate}

\item Our morphological analysis of the deep \textit{Chandra} data
  confirms the presence of deep, inner X-ray cavities discovered in
  the previous snapshot observations. Using the surface brightness
  profile, we estimate the diameters of the cavities to be $\sim$ 20
  kpc.
  We also note the presence of edges surrounding the central region of
  the cluster, which could be the signature of weak shocks that
  supplement the energy released by the cavity system.

\item We find an enhanced iron abundance in a region that is extended
  in the direction of the cavities, as predicted by numerical models.
  This suggests that the AGN outbursts that inflate the cavities also
  lift metal enriched gas outward in the direction of the radio jets.
  Therefore, the AGN in the core contributes to lift the metals with
  its outburst which drives the cavity inflation.  Due to the limited
  statistics we are unable to create a metallicity map of the cluster
  which would be necessary to perform a more detailed study.

\item The power of the cavity system is estimated to be P$_{cav}$ =
  2.5 $\times$ 10$^{45}$ erg s$^{-1}$ (Table \ref{tab:cavity
    energetics}), consistent with the value estimated by
  \cite{Cavagnolo_2011}.
  Comparing to the X-ray luminosity of the gas within the cooling
  radius ($r_{\rm cool} = 109 \, {\rm kpc}$), estimated using
  deprojected profiles as L$_{X}$ $=$ 1.33 $\times$ 10$^{45}$ erg
  s$^{-1}$, indicates that the mechanical luminosity of the AGN
  outburst is large enough to balance the radiative losses. The cavity
  power gives a lower limit on the true power of the AGN, as the
  possible presence of further cavities and weak shocks would
  contribute to it thus providing a more complete estimate of the jet
  power.

\item We find indication that the cavities have higher temperature
  than the surrounding ICM. A support for this consideration comes
  both from the projected spectral analysis and from the
  softness ratio method described in Section \ref{subsec:metal}.
  However, this may only be a projection effect, as we find that the
    cavities are consistent with being empty of X-ray emitting gas.

\item Our new \textit{VLA} data detect a 1.4 GHz diffuse radio
  emission with size comparable to the cooling region (Figure
  \ref{fig:chandra_radio}). We show how this emission can be
  classified as a radio mini-halo, following the trend
  P$_{radio}$-P$_{CF}$ predicted by the reacceleration model (Section
  \ref{minihalo.sec}).
\end{enumerate}

\acknowledgments

AD is member of the International Max Planck Research School (IMPRS) for
Astronomy and Astrophysics at the Universities of Bonn and Cologne. Significant
part of this work was done by AD as a Master student at the University of
Bologna. AD thanks the Harvard-Smithsonian Center for Astrophysics for the
hospitality during the finalization of this work and acknowledges support by the
Chandra grant GO0-11136X.
MG acknowledges support by grants ASI-INAF I/023/05/0 and I/088/06/0 and by
Chandra grants GO0-11003X and GO0-11136X.
BRM acknowledges support from the Natural Sciences and Engineering Research
Council of Canada.

\bibliographystyle{apj}
\bibliography{bibliography}

\end{document}